\documentclass[twocolumn,english,rmp]{revtex4-1}
\usepackage[T1]{fontenc}
\usepackage[utf8]{inputenc}
\usepackage{color}
\usepackage{babel}
\usepackage{graphicx}
\usepackage{amsfonts}
\usepackage{amsmath}
\usepackage{natbib}
\usepackage{hyperref}

\begin{document}

\title{Machine learning and the physical sciences}

\author{Giuseppe Carleo}
\affiliation{Center for Computational Quantum Physics, Flatiron Institute,\\ 162
5th Avenue, New York, NY 10010, USA}
\email[Corresponding author: ]{gcarleo@flatironinstitute.org}

\author{Ignacio Cirac}
\affiliation{Max-Planck-Institut fur Quantenoptik,\\ Hans-Kopfermann-Straße 1, D-85748 Garching, Germany}

\author{Kyle Cranmer}
\affiliation{Center for Cosmology and Particle Physics, Center of Data Science, \\ New York University,
726 Broadway, New York, NY 10003, USA}

\author{Laurent Daudet}
\affiliation{LightOn, 2 rue de la Bourse, F-75002 Paris, France}

\author{Maria Schuld}
\affiliation{University of KwaZulu-Natal, Durban 4000, South Africa \\
National Institute for Theoretical Physics, KwaZulu-Natal, Durban 4000, South Africa,\\
and Xanadu Quantum Computing, 777 Bay Street, M5B 2H7 Toronto, Canada}

\author{Naftali Tishby}
\affiliation{The Hebrew University of Jerusalem, Edmond Safra Campus, Jerusalem 91904, Israel}

\author{Leslie Vogt-Maranto}
 \affiliation{Department of Chemistry, New York University, New York, NY 10003, USA}

\author{Lenka Zdeborov\'a}
\affiliation{Institut de physique th\'eorique, Universit\'e Paris Saclay, CNRS, CEA,\\ F-91191 Gif-sur-Yvette, France}
\email[Corresponding author: ]{lenka.zdeborova@cea.fr}

\begin{abstract}
Machine learning encompasses a broad range of algorithms and modeling tools used for a vast array of data processing tasks, which has entered most scientific disciplines in recent years.
We review in a selective way the recent research on the interface between machine learning and physical sciences. This includes conceptual developments in machine learning (ML) motivated by physical insights, applications of machine learning techniques to several domains in physics, and cross-fertilization between the two fields.
After giving basic notion of machine learning methods and principles, we describe examples of how statistical physics is used to understand methods in ML. We then move to describe applications of ML methods in particle physics and cosmology, quantum many body physics, quantum computing, and chemical and material physics. We also highlight research and development into novel computing architectures aimed at accelerating ML. In each of the sections we describe recent successes as well as domain-specific methodology and challenges.
\tableofcontents{}
\end{abstract}
\maketitle

\section{Introduction}
The past decade has seen a prodigious rise of machine-learning (ML) based
techniques, impacting many areas in industry including autonomous driving, health-care, finance, manufacturing, energy harvesting, and more.
ML is largely perceived as one of the main disruptive technologies of our ages,
as much as computers have been in the 1980's and 1990's.
The general goal of ML is to recognize patterns in data, which inform the way unseen problems are treated. For example, in a highly complex system such as a self-driving car, vast amounts of data coming from sensors have to be turned into decisions of how to control the car by a computer that has ``learned'' to recognize the pattern of ``danger''.

The success of ML in recent times has been
marked at first by significant improvements on some existing technologies, for example in the field of image recognition.
To a large extent, these advances constituted the first demonstrations of the impact
that ML
methods can have in specialized tasks.
More recently, applications traditionally inaccessible to automated software have
been successfully enabled, in particular by deep learning technology. The demonstration of reinforcement learning techniques in game playing, for example, has
had a deep impact in the perception that the whole field was moving a
step closer to what expected from a general artificial intelligence.


In parallel to the rise of ML techniques in industrial applications, scientists have increasingly become interested
in the potential of ML for fundamental research, and physics is no exception.
To some extent, this is not too surprising, since both ML and physics share some of their methods as well as goals.
The two disciplines are both concerned about the process of gathering and analyzing data to design models that can predict the behaviour of complex systems.
However, the fields prominently differ in the way their fundamental goals are realized.
On the one hand, physicists want to understand the mechanisms of Nature, and are proud of using their own knowledge, intelligence and intuition to inform their models. On the other hand, machine learning mostly does the opposite: models are agnostic and the machine provides the 'intelligence' by extracting it from data. Although often powerful, the resulting models are notoriously known to be as opaque to our understanding as the data patterns themselves. Machine learning tools in physics are therefore welcomed enthusiastically by some, while being eyed with suspicions by others. What is difficult to deny is that they produce surprisingly good results in some cases.

In this review, we attempt at providing a coherent selected account of the diverse intersections of ML with physics.
Specifically, we look at an ample spectrum of fields (ranging from statistical and quantum physics to high energy and cosmology) where ML recently made a prominent appearance, and discuss potential applications and challenges of `intelligent' data mining techniques in the different contexts.
We start this review with the field of statistical physics in Section \ref{sec:stat_phys} where the interaction with machine learning has a long history,
drawing on methods in physics to provide better understanding of problems in machine learning. We then turn the wheel in the other direction of using machine learning for physics. Section \ref{sec:particle} treats progress in the fields of high-energy physics and cosmology,  Section \ref{sec:many_body} reviews how ML ideas are helping to understand the mysteries of many-body quantum systems,  Section \ref{sec:quantum} briefly explore the promises of machine learning within quantum computations, and in Section \ref{sec:chem_mat} we highlight some of the amazing advances in computational chemistry and materials design due to ML applications. In Section \ref{sec:instruments} we discuss some advances in instrumentation leading potentially to hardware adapted to perform machine learning tasks. We conclude with an outlook in Section \ref{sec:outlook}.


\subsection{Concepts in machine learning}

For the purpose of this review we will briefly explain some fundamental terms and concepts used in machine learning. For further reading, we recommend a few resources, some of which have been targeted especially for a physics audience. For a historical overview of the development of the field we recommend Refs.~\cite{lecun2015deep,DBLP:journals/corr/Schmidhuber14}. An excellent recent introduction to machine learning
for physicists is Ref. \cite{mehta2018high}, which includes notebooks
with practical demonstrations. A very useful online resource is Florian Marquardt's course ``Machine learning for physicists''\footnote{ See \url{https://machine-learning-for-physicists.org/}.}. Useful textbooks written by machine learning researchers are Christopher Bishop's standard textbook \citep{bishop2006pattern}, as well as \cite{goodfellow2016deep} which focuses on the theory and foundations of deep learning and covers many aspects of current-day research. A variety of online tutorials and lectures is useful to get a basic overview and get started on the topic.

To learn about the theoretical progress made in statistical
physics of neural networks in the 1980s-1990s we recommend the rather accessible book
Statistical Mechanics of Learning \cite{engel2001statistical}. For learning details of the replica method and its use in
computer science, information theory and machine learning we would
recommend the book of Nishimori \cite{nishimori2001statistical}. For the more recent
statistical physics methodology the textbook of M{\'e}zard and
Montanari is an excellent reference \cite{mezard2009information}.


To get a basic idea of the type of problems that machine learning is
able to tackle it is useful to defined three large classes of learning
problems: Supervised learning, unsupervised learning and reinforcement
learning. This will also allow us to state the basic terminology,
building basic equipment to expose some of the basic tools of
machine learning.

\subsubsection{Supervised learning and neural networks}

 In supervised learning we are given a set of $n$ samples of data, let
 us denote one such sample $X_\mu \in  {\mathbb R}^p$,  with $\mu = 1,
 \dots, n$. To have something concrete in mind each $X_\mu$ could be
 for instance a black-and-white photograph of an animal, and $p$ the
 number of pixels. For each sample $X_\mu$ we are further given a
 {\it label} $y_\mu \in {\mathbb R}^d$, most commonly $d=1$.
 The label could encode for instance the species of the animal on the photograph.
 The goal of supervised learning is to find a function $f$ so that when a new sample $X_{\rm
 new}$ is presented without its label, then the output of the function $f(X_{\rm
 new})$ approximates well the label. The data set $\{X_\mu,y_\mu\}_{\mu = 1,\dots,n}$ is called the {\it
  training set}. In order to test the resulting function $f$ one
usually splits the available data samples into the training set used
to learn the function and a {\it test set} to evaluate the
performance.

Let us now describe the {\it training} procedure most commonly used to find a
suitable function $f$. Most commonly the function is
expressed in terms of a set of parameters, called {\it weights} $w \in {\mathbb R}^k$,
leading to $f_w$. One then constructs a so-called {\it loss function} ${\cal
L}[f_w(X_\mu),y_\mu]$ for each sample $\mu$, with the idea of
this loss being small when $f_w(X_\mu)$ and $y_\mu$ are close, and
vice versa. The average of the loss over the training set is then
called the {\it empirical risk} ${\cal R}(f_w) = \sum_{\mu =1}^n {\cal
L}[f_w(X_\mu),y_\mu] / n$.

During the training procedure the weights $w$ are being adjusted
in order to minimize the empirical risk. The {\it training error}
measures how well is such a minimization achieved. A notion of error
that is the most important is the {\it generalization error}, related to
the performance on predicting labels $y_{\rm new}$ for data samples
$X_{\rm new}$ that were not seen in the training set. In applications,
it is common practice to build the test set by randomly picking
a fraction of the available data,
and perform the training using the remaining fraction as a training set.
We note that in a part of the
literature the {\it generalization error} is the difference between the
performance of the test set and the one on the training set.

The algorithms most commonly used to minimize the empirical risk function over the weights are based on {\it gradient descent} with respect to the weights
$w$. This means that the weights are iteratively adjusted in the
direction of the gradient of the empirical risk
\begin{equation}
     w^{t+1} = w^{t} - \gamma \nabla_w {\cal R}(f_w).
\end{equation}
 The rate $\gamma$ at which this is performed is called the {\it learning
   rate}. A very commonly used and successful variant of the gradient descent is
the {\it stochastic gradient descent} (SGD) where the full empirical
risk function ${\cal R}$ is replaced by the contribution of
just a few of the samples. This subset of samples is called {\it
  mini-batch} and can be as small as a single sample. In physics terms,
the SGD algorithm is often compared to the Langevin dynamics at finite
temperature. Langevin dynamics at zero temperature is the
gradient descent. Positive temperature introduces a thermal noise that
is in certain ways similar to the noise arising in SGD, but different
in others. There are many variants of the SGD algorithm used in
practice. The initialization of the weights can change performance in
practice, as can the choice of the learning rate and a variety of so-called
regularization terms, such as {\it weight decay} that is penalizing
weights that tend to converge to large absolute values. The choice of
the right version of the algorithm is important, there are many
heuristic rules of thumb, and certainly more theoretical insight into
the question would be desirable.

One typical example of a task in supervised learning is {\it classification},
that is when the labels $y_\mu$ take values in a discrete set and the
so-called {\it accuracy} is then measured as the fraction of times the
learned function classifies the data point correctly. Another example
is {\it regression} where the goal is to learn a real-valued function, and
the accuracy is typically measured in terms of the {\it mean-squared
error} between the true labels and their learned estimates. Other
examples would be sequence-to-sequence learning where both the input
and the label are vectors of dimension larger than one.

There are many methods of supervised learning and many variants of
each. One of the most basic supervised learning method is the widely
known and used {\it linear regression}, where the function $f_w(X)$ is
parameterized in the form $f_w(X_\mu) = X_\mu w$, with $w \in {\mathbb R}^p$. When the data live in high
dimensional space and the number of samples is not much larger than
the dimension, it is indispensable to use regularized form of
linear regression called {\it ridge regression} or {\it Tikhonov regularization}. The ridge regression
is formally equivalent to assuming that the weights $w$ have a
Gaussian prior. A generalized form of linear regression, with parameterization
$f_w(X_\mu) = g(X_\mu w)$, where $g$ is some {\it output channel}
function, is also often used and its properties are described in
section \ref{sec:GLM}. Another popular way of regularization is based on separating the example n a classification task so that they the separate categories are divided by a clear gap that is as wide as possible. This idea stands behind the definition of so-called {\it support vector machine} method.

A rather powerful non-parametric generalization of the ridge
regression is {\it kernel ridge regression}. Kernel ridge regression is closely related to {\it Gaussian process regression}. The support vector machine method is often combined with a kernel method, and as such is still the state-of-the-art method in many applications, especially when the number of available samples is not very large.

Another classical supervised learning method is based on so-called {\it decision trees}. The decision tree is used to go from observations about a data sample (represented in the branches) to conclusions about the item's target value (represented in the leaves). The best known application of decision trees in physical science is in data analysis of particle accelerators, as discussed in Sec.~\ref{sec:Class_part}.

The supervised learning method that stands behind the machine learning
revolution of the past decade are multi-layer {\it feed-forward neural
networks} (FFNN) also sometimes called {\it multi-layer perceptrons}. This is also a very relevant method for the purpose of
this review and we shall describe it briefly here. In {\it $L$-layer fully
connected neural networks} the function $f_w(X_\mu)$ is parameterized
as follows
\begin{equation}
      f_w(X_\mu) =  g^{(L)}(W^{(L)}  \dots g^{(2)} (W^{(2)} g^{(1)}(W^{(1)} X_\mu ))), \label{eq:ffnn}
\end{equation}
where $w=\{W^{(1)},\dots,W^{(L)}\}_{i=1,\dots,L}$, and $W^{(i)} \in {\mathbb R}^{r_i \times r_{i-1}}$ with $r_{0} = p$
and $r_{L} = d$, are the matrices of weights, and $r_i$ for $1 \le i \le
L - 1$ is called the
{\it width} of the $i-{\rm th}$ {\it hidden layer}. The functions $g^{(i)}$, $1 \le i \le
L$, are the so-called {\it activation functions}, and act component-wise on
vectors. We note that the input in the activation functions are often slightly more generic affine transforms of the output of the previous layer that simply matrix multiplications, including e.g. biases. The number of layers $L$ is called the network's {\it depth}. Neural networks with depth larger than some small integer are called {\it deep neural networks}. Subsequently machine learning based on deep neural networks is called {\it deep learning}.

The theory of neural
networks tells us that without hidden layers ($L=1$, corresponding to
the generalized linear regression) the set of functions that can be
approximated this way is very limited \cite{minsky69perceptrons}. On the other hand
already with one hidden layer, $L=2$, that is wide enough, i.e. $r_1$
large enough, and where the function $g^{(1)}$ is non-linear, a very general class of functions can be well
approximated in principle \cite{cybenko1989approximation}. These theories, however, do not
tell us what is the optimal set of parameters (the activation
functions, the widths of the layers and the depth) in order
for the learning of $W^{(1)}, \dots, W^{(L)}$ to be tractable
efficiently. We know from empirical success of the past decade that
many tasks of interest are tractable with deep neural network
using the gradient descent or the SGD algorithms.
In deep neural networks the derivatives with respect
to the weights are computed using the chain rule leading to the
celebrated {\it back-propagation algorithm} that takes care of efficiently
scheduling the operations required to compute all the gradients \cite{goodfellow2016deep}.

A very important and powerful variant of (deep) feed-forward neural networks
are the so-called {\it convolutional neural networks} \cite{goodfellow2016deep} where the input
into each of the hidden units is obtained via a filter applied to
a small part of the input space. The filter is then shifted to
different positions corresponding to different hidden
units. Convolutional neural networks implement invariance to
translation and are in particular suitable for analysis of
images. Compared to the fully connected neural networks each layer of
convolutional neural network has much smaller number of parameters,
which is in practice advantageous for the learning algorithms. There are many types and variances of convolutional neural networks, among them we will mention the Residual neural networks (ResNets) use shortcuts to jump over some layers.

Next to feed-forward neural networks there are the so-called {\it recurrent neural networks} (RNN) in which the outputs of units feeds back at the input in the next time step. In RNNs the result is thus given by the set of weights, but also by the whole temporal sequence of states. Due to their intrinsically dynamical nature, RNN are particularly suitable for learning for temporal data sets, such as speech, language, and time series. Again there are many types and variants on RNNs, but the ones that caused the most excitement in the past decade are arguably the long short-term memory (LSTM) networks \cite{hochreiter1997long}. LSTMs and their deep variants are the state-of-the-art in tasks such as speech processing, music compositions, and natural language processing.

\subsubsection{Unsupervised learning and generative modelling}

Unsupervised learning is a class of learning problems where input data
are obtained as in supervised learning, but no labels
are available. The goal of learning here is to recover some
underlying --and possibly non-trivial-- structure in the dataset. A typical example of unsupervised learning is data
clustering where data points are assigned into groups in such a way
that every group has some common properties.

In unsupervised learning, one often seeks a probability
distribution that generates samples that are statistically similar
to the observed data samples, this is often referred to as {\it generative modelling}. In some cases this probability
distribution is written in an explicit form and explicitly or implicitly parameterized. Generative models internally contain {\it latent variables} as the source of randomness.
When the number of latent variables is much smaller
than the dimensionality of the data we speak of
{\it dimensionality reduction}.  One path towards unsupervised learning is
to search values of the latent variables that maximize the likelihood of the observed data.

In a range of applications the likelihood associated to the observed data is not known or computing it is itself intractable. In such cases, some of the generative models discussed below offer on alternative likelihood-free path. In Section \ref{sec:likelihood_free} we will also discuss the so-called {\it ABC method} that is a type of likelihood-free inference and turns out to be very useful in many contexts arising in physical sciences.

Basic methods of unsupervised learning include {\it principal component
analysis} and its variants. We will cover some theoretical insights
into these method that were obtained using physics in section
\ref{sec:PCA}. A physically very appealing methods for unsupervised learning
are the so-called {\it Boltzmann machines} (BM).
A BM is basically an inverse Ising model
where the data samples are seen as samples from a Boltzmann
distribution of a pair-wise interacting Ising model. The goal is to
learn the values of the interactions and magnetic fields so that the
likelihood (probability in the Boltzmann measure) of the observed data
is large. A {\it restricted Boltzmann machine} (RBM) is a particular case of BM
where two kinds of variables -- visible units, that see the input data, and
hidden units, interact through effective couplings.
The interactions are in this case only between visible and hidden
units and are again adjusted in order for the likelihood of the
observed data to be large. Given the appealing interpretation
in terms of physical models, applications of BMs and RBMs are widespread
in several physics domains, as discussed e.g. in section \ref{sec:NNQS}.

A very neat idea to perform unsupervised learning yet being able to use
all the methods and algorithms developed for supervised learning are
{\it auto-encoders}. An autoencoder is a feed-forward neural network
that has the input data on the input, but also on the output. It aims
to reproduce the data while typically going trough a bottleneck, in
the sense that some of the intermediate layers have very small width
compared to the dimensionality of the data. The idea is then that
autoencoder is aiming to find a succinct representation of the data
that still keeps the salient features of each of the samples.  Variational autoencoders (VAE) \cite{kingma2013auto,2014arXiv1401.4082J} combine variational inference and autoencoders to provide a deep generative model for the data, which can be trained in an unsupervised fashion.

A further approach to unsupervised learning worth mentioning here, are {\it adversarial
  generative networks} (GANs) \cite{goodfellow2014generative}.
  GANs have attracted substantial attentions in the past years,
  and constitute another fruitful way to take advantage of the progresses made for supervised learning to do
unsupervised one. GANs typical use two feed-forward neural networks,
one called the {\it generator} and another called the {\it
  discriminator}. The generator network is used to generate outputs from
random inputs, and is designed so that the outputs look like the observed samples. The discriminator network  is used to discriminate
between true data samples and samples generated by the generator
network. The discriminator is aiming at best possible accuracy in this
classification task, whereas the generator network is adjusted to make
the accuracy of the discriminator the smallest possible. GANs
currently are the state-of-the art system for many applications in
image processing.

Other interesting methods to model distributions include {\it normalizing flows} and {\it autoregressive models} with the advantage of having tractable likelihood so that they can be trained via maximum likelihood~\cite{larochelle2011neural,uria_neural_2016,papamakarios2017masked}.


Hybrids between supervised learning and unsupervised learning that are
important in application include {\it semi-supervised learning} where only
some labels are available, or {\it active learning} where labels can
be acquired for a selected set of data points at a certain cost.

\subsubsection{Reinforcement learning}

Reinforcement learning \cite{sutton2018reinforcement} is an area of machine learning where an
(artificial) agent takes actions in an environment with the goal of
maximizing a reward. The action changes the state of the environment in some way and the agent typically observes some information about the state of the environment and the corresponding reward. Based on those observations the agent decides on the next action, refining the strategies of which action to choose in order to maximize the resulting reward. This type of learning is designed for cases where the only way to learn about the properties of the environment is to interact with it. A key concept in reinforcement learning it the trade-off between exploitation of good strategies found so far, and exploration in order to find yet better strategies. We should also note that reinforcement learning is intimately related to the field of theory of control, especially {\it optimal control theory}.

One of the main types of reinforcement learning applied in many works is the so-called {\it Q-learning}. Q-learning is based on a value matrix $Q$ that assigns quality of a given action when the environment is in a given state. This value function $Q$ is then iteratively refined. In recent advanced applications of Q-learning the set of states and action is so large that it is impossible to even store the whole matrix~$Q$. In those cases deep feed-forward neural networks are used to represent the function in a succinct manner. This gives rise to {\it deep Q-learning}.

Most well-known recent examples of the success of reinforcement learning is the computer program AlphaGo and AlphaGo Zero that for a first time in history reached super-human performance in the traditional board game of Go. Another well known use of reinforcement learning is locomotion of robots.














\section{Statistical Physics }
\label{sec:stat_phys}

\subsection{Historical note}

While machine learning as a wide-spread tool for physics research is a relatively new phenomenon, cross-fertilization between the two disciplines dates back much further. Especially statistical physicists made important contributions to our theoretical understanding of learning (as the term ``statistical'' unmistakably suggests).

The connection between statistical mechanics and learning theory
started when statistical learning from examples took over the logic
and rule based AI, in the mid 1980s. Two seminal papers marked this
transformation, Valiant's theory of the learnable \cite{valiant1984theory},
which opened the way for rigorous statistical learning in AI, and
Hopfield's neural network model of associative memory \cite{hopfield1982neural}, which sparked the rich application
of concepts from spin glass theory to neural networks models. This was
marked by the memory capacity calculation of the Hopfield model by
Amit, Gutfreund, and Sompolinsky \cite{amit1985spin} and following
works. A much tighter application to learning models was made by the
seminal work of Elizabeth Gardner who applied the replica trick \cite{gardner1987maximum,gardner1988space} to calculate volumes in the weights space for simple feed-forward neural networks, for both supervised and unsupervised learning models.

Gardner's method enabled to explicitly calculate {\it learning
curves}, i.e. the typical training and generalization errors as a function of the number of training examples, for very specific one and two-layer
neural networks \cite{gyorgyitishby1990learning,sompolinsky1990learning,seung1992statistical}. These analytic statistical physics
calculations demonstrated that the learning dynamics can exhibit much
richer behavior than predicted by the worse-case distribution
free PAC bounds (PAC stands for {\it provably
    approximately correct}) \cite{valiant1984theory}. In particular, learning can
exhibit phase transitions from poor to good
generalization \cite{gyorgyi1990first}. This rich learning dynamics and curves can
appear in many machine learning problems, as was
shown in various models, see e.g. more recent review \cite{zdeborova2016statistical}. The statistical physics of learning reached
its peak in the early 1990s, but had rather minor influence on
machine-learning practitioners and theorists, who were focused on
general input-distribution-independent generalization bounds, characterized by e.g. the Vapnik-Chervonenkis dimension or the Rademacher complexity of hypothesis classes.

\subsection{Theoretical puzzles in deep learning}

Machine learning in the new millennium was marked by much larger scale learning problems, in input/pattern
sizes which moved from hundreds to millions in dimensionality, in
training data sizes, and in number of adjustable parameters. This was
dramatically demonstrated by the return of large scale feed-forward
neural network models, with many more hidden layers, known as deep
neural networks. These deep neural networks were essentially the same
feed-forward convolution neural networks proposed already in the 80s.
But somehow with the much larger scale inputs and big and clean training data
(and a few more tricks and hacks), these networks started to beat the state-of-the-art in
many different pattern recognition and other machine learning
competitions, from roughly 2010 and
on. The amazing performance of deep learning, trained with the same
old {\it stochastic gradient descent} (SGD) error-back-propagation algorithm, took everyone by surprise.

One of the puzzles is that the existing learning theory (based on the worst-case PAC-like generalization bounds) is unable to explain this phenomenal success. The existing theory does not predict why deep networks, where the number/dimension of adjustable parameters/weights is way higher than the number of training samples, have good generalization properties. This lack of theory was coined in now a classical article \cite{zhang2016understanding}, where the authors show numerically that state-of-the-art neural networks used for classification are able to classify perfectly randomly generated labels. In such a case existing learning theory does not provide any useful bound on the generalization error. Yet in practice we observe good generalization of the same deep neural networks when trained on the true labels.

Continuing with the open question, we do not have good understanding of which learning problems are computationally tractable. This is particularly important since from the point of view of computational complexity theory, most of the learning problems we encounter are NP-hard in the worst case. Another open question that is central to current deep learning concern the choice of hyper-parameters and architectures that is so far guided by a lot of trial-and-error combined by impressive experience of the researchers. At the same time as applications of ML are spreading into many domains, the field calls for more systematic and theory-based approaches. In current deep-learning, basic questions such as what is the minimal number of samples we need in order to be able to learn a given task with a good precision is entirely open.

At the same time the current literature on deep learning is flourishing with interesting numerical observations and experiments that call for explanation. For a physics audience the situation could perhaps be compared to the state-of-the-art in fundamental small-scale physics just before quantum mechanics was developed. The field was full of unexplained experiments that were evading existing theoretical understanding. This clearly is the perfect time for some of physics ideas to study neural networks to resurrect and revisit some of the current questions and directions in machine learning.


Given the long history of works done on neural networks in
statistical physics, we will not aim at a complete review of this direction of research. We will focus in a selective way on
recent contributions originating in physics that, in our opinion, are having important impact in current theory of learning and machine learning. For the purpose of this review we are also putting aside a large volume of work done in statistical physics on recurrent neural networks with biological applications in mind.

\subsection{Statistical physics of unsupervised learning}

\subsubsection{Contributions to understanding basic unsupervised methods}
\label{sec:PCA}

One of the most basic tools of unsupervised learning across the sciences
are methods based on low-rank decomposition of the observed data matrix. Data
clustering, principal component analysis (PCA), independent component
analysis (ICA), matrix completion,  and other methods are examples in this class.

In mathematical language the low-rank matrix decomposition
problem is stated as follows: We observe $n$ samples of
$p$-dimensional data $x_i \in {\mathbb R}^p$, $i=1,\dots,n$. Denoting
$X$ the $n\times p$ matrix of data, the idea
underlying low-rank decomposition methods assumes that $X$ (or some component-wise
function of $X$) can be written as a noisy version of a rank $r$
matrix where $r \ll p; r\ll n$, i.e. the rank is much lower that the dimensionality
and the number of samples, therefore the name {\it low-rank}.
A particularly challenging, yet relevant and interesting regime, is when the
dimensionality $p$ is comparable to the number of samples $n$, and
when the level of noise is large in such a way that perfect
estimation of the signal is not possible. It turns out that the low-rank matrix estimation in
the high-dimensional noisy regime can be modelled as a statistical
physics model of a {\it spin glass} with $r$-dimensional vector variables
and a special {\it planted} configuration to be found.

Concretely, this
model can be defined in the teacher-student scenario in which
the teacher generates $r$-dimensional latent variables $u^*_i \in {\mathbb
  R}^r$, $i=1,\dots,n$, taken from a given probability distribution $P_u(u_i^*)$, and
$r$-dimensional latent variables $v^*_j \in {\mathbb
  R}^r$, $j=1,\dots,p$, taken from a given probability distribution $P_v(v_i^*)$. Then the teacher generates components of the
data matrix $X$ from some given conditional probability distribution $P_{\rm
  out}(X_{ij} | u^*_i \cdot v^*_j)$. The goal of the student is then
to recover the latent variables $u^*$ and $v^*$ as precisely as
possible from the knowledge of $X$, and the distributions $P_{\rm
  out}$, $P_u$, $P_v$.

Spin glass theory can be used to obtain rather complete understanding of this
teacher-student model for low-rank matrix estimation
in the limit $p,n\to \infty$, $n/p=\alpha = \Omega(1), r= \Omega(1)$.
One can compute with the replica method what is the information-theoretically best error in estimation of
$u^*$, and $v^*$ the student can possibly achieve, as done decades ago for some special choices of $r$, $P_{\rm out}$, $P_u$ and $P_v$ in \cite{biehl1993statistical,watkin1994optimal,barkai1994statistical}. The
importance of these early works in physics is acknowledged in some of
the landmark papers on the subject in statistics, see
e.g. \cite{johnstone2009consistency}. However, the lack of mathematical
rigor and limited understanding of algorithmic tractability caused the impact of these
works in machine learning and statistics to remain limited.

A resurrection of interest in statistical physics approach to low-rank matrix
decompositions came with the study of the stochastic block model
for detection of clusters/communities in sparse networks. The
problem of community detection was studied heuristically and
algorithmically extensively in statistical physics, for a review see
\cite{fortunato2010community}. However, the exact solution and understanding of algorithmic
limitations in the stochastic block model came from the spin glass theory in
\cite{decelle2011inference,decelle2011asymptotic}. These works
computed (non-rigorously) the asymptotically optimal performance and delimited sharply
regions of parameters where this performance is reached by the belief
propagation (BP) algorithm \cite{yedidia2003understanding}. Second order phase transitions appearing in the
model separate a phase where clustering cannot be performed better
than by random guessing, from a region where it can be done
efficiently with BP. First order phase transitions and
one of their spinodal lines then separate regions where clustering is
impossible, possible but not doable with the BP algorithm, and
easy with the BP algorithm. Refs.~\cite{decelle2011inference,decelle2011asymptotic} also conjectured that when the BP algorithm is not able to reach the optimal performance on large
instances of the model, then no other polynomial algorithm will. These
works attracted a large amount of follow-up work in mathematics,
statistics, machine learning and computer science communities.

The statistical physics understanding of the
stochastic block model and the conjecture about belief propagation
algorithm being optimal among all polynomial ones inspired the discovery
of a new class of spectral algorithms for sparse data (i.e. when the
matrix $X$ is sparse) \cite{krzakala2013spectral}. Spectral algorithms are basic tools in data
analysis \cite{ng2002spectral,von2007tutorial}, based on the singular
value decomposition of the matrix $X$ or functions of $X$. Yet for
sparse matrices $X$, the spectrum is known to have leading singular
values with localized singular vectors unrelated to the latent
underlying structure. A more robust spectral method is obtained by
linearizing the belief propagation, thus obtaining a so-called
non-backtracking matrix \cite{krzakala2013spectral}.  A variant on
this spectral method based on algorithmic interpretation of the
Hessian of the Bethe free energy also originated in physics \cite{saade2014spectral}.


This line of statistical-physics inspired research is merging into the mainstream in statistics and machine learning. This
is largely thanks to recent progress in: (a) our understanding of
algorithmic limitations, due to the analysis of approximate message
passing (AMP) algorithms \cite{rangan2012iterative,matsushita2013low,deshpande2014information,bolthausen2014iterative,javanmard2013state} for low-rank matrix estimation that is a
generalization of the Thouless-Anderson-Palmer equations
\cite{thouless1977solution} well known in the physics literature on
spin glasses. And (b) progress in proving many of the corresponding results
in a mathematically rigorous way. Some of the influential papers in
this direction (related to low-rank matrix estimation) are
\cite{deshpande2014information,barbier2016mutual,lelarge2016fundamental,coja2018information}
for the proof of the replica formula for the information-theoretically
optimal performance.

\subsubsection{Restricted Boltzmann machines}

Boltzmann machines and in particular restricted Boltzmann machines are
another method for unsupervised learning often used in machine
learning. As apparent from the very name of the method, it had strong
relation with statistical physics. Indeed the Boltzmann machine is
often called the inverse Ising model in the physics literature and
used extensively a range of area, for a recent review on the physics
of Boltzmann machines see \cite{nguyen2017inverse}.

Concerning restricted Boltzmann machines, there are number of studies in
physics clarifying how these machines work and what structures can they
learn.
Model of random restricted Boltzmann machine, where the weights are imposed to be
random and sparse, and not learned, is studied in \cite{tubiana2017emergence,cocco2018statistical}. Rather
remarkably for a range of potentials on the hidden unit this work
unveiled that even the single layer RBM is able to represent
compositional structure. Insights from this work were more recently
used to model protein families from their sequence information
\cite{tubiana2018learning}.

Analytical study of the learning process in RBM, that is most commonly
done using the {\it contrastive divergence} algorithm based on Gibbs
sampling \cite{hinton2002training}, is very
challenging. First steps were studied in \cite{decelle2017spectral} at
the beginning of the learning process where the dynamics can be
linearized. Another interesting direction coming from statistical
physics is to replace the Gibbs sampling in the contrastive divergence
training algorithm by the Thouless-Anderson-Palmer equations
\cite{thouless1977solution}. This has been done in
\cite{gabrie2015training,tramel2018deterministic} where such training was shown to be competitive and applications of the approach were discussed.
RBM with random weights and their relation to the Hopfield model was
clarified in \cite{mezard2017mean,barra2018phase}.

\subsubsection{Modern unsupervised and generative modelling}

The dawn of deep learning brought an exciting innovations into
unsupervised and generative-models learning. A physics friendly
overview of some classical and more recent concepts is e.g. \cite{wang2018generative}.

Auto-encoders with linear activation functions are closely related to PCA.
Variational autoencoders (VAE) \cite{kingma2013auto,2014arXiv1401.4082J} are variants much
closer to a physicist mind set where the autoencoder is represented
via a graphical model, and in trained using a prior on the latent variables and
variational inference. VAE with a single hidden
layer is closely related to other widely used techniques in signal
processing such as dictionary learning and sparse coding. Dictionary
learning problem has been studied with statistical physics techniques in
\cite{sakata2013statistical,krzakala2013phase,kabashima2016phase}.

Generative adversarial networks (GANs) -- a powerful set of ideas emerged
with the work of \cite{goodfellow2014generative} aiming to generate samples
(e.g. images of hotel bedrooms) that are of the same type as those in the training
set.
Physics-inspired studies of GANs are starting to
appear, e.g. the work  on a solvable model of GANs by
\cite{wang2018solvable} is a intriguing generalization of the earlier
statistical physics works on online learning in perceptrons.

We also want to point the readers attention to autoregressive
generative models \cite{larochelle2011neural,uria_neural_2016,papamakarios2017masked}. The main interest in autoregressive models stems from the fact that they are a family of explicit probabilistic models, for which direct and unbiased sampling is possible.  Applications of these models have been realized for both statistical \cite{wu2018solving} and quantum physics problems \cite{sharir_deep_2019}.


\subsection{Statistical physics of supervised learning}

\subsubsection{Perceptron and GLMs}
\label{sec:GLM}

The arguably most basic method of supervised learning is linear regression
where one aims to find a vector of coefficients $w$ so that its scalar product
with the data point $X_i  w$ corresponds to the observed predicate
$y$. This is most often solved by the {\it least squares method} where $|| y -
X w||^2_2$ is minimized over $w$. In the Bayesian language, the least
squares method corresponds to assuming Gaussian additive noise $\xi$ so
that $y_i =X_i w + \xi_i $. In high dimensional setting it is
almost always indispensable to use regularization of the weights. The most
common ridge regularization corresponds in the Bayesian interpretation
to Gaussian prior on the weights. This probabilistic thinking can be
generalized by assuming a general prior $P_W(\cdot)$ and a generic
noise represented by a conditional probability distribution $P_{\rm
  out}(y_i| X_i w)$. The resulting model is called {\it
  generalized linear regression} or {\it generalized linear model} (GLM).
Many other problems of interest in data analysis and learning can be
represented as GLM. For instance sparse regression simply requires
that $P_W$ has large weight on zero, for the perceptron with threshold
$\kappa$ the output has
a special form $P_{\rm  out}(y| z) = {\mathbb I}(z>\kappa) \delta(y-1)
+  {\mathbb I}(z\le \kappa) \delta(y+1)$. In the language of neural
networks, the GLM represents a single layer (no hidden variables) fully connected
feed-forward network.

For generic noise/activation channel $P_{\rm  out}$  traditional theories in statistics are not readily applicable to the
regime of very limited data where both the dimension $p$ and the number of
samples $n$ grow large, while their ratio $n/p=\alpha$ remains
fixed. Basic questions such as: how does the best achievable
generalization error depend on the number of samples, remain open. Yet
this regime and related questions are of great interest and
understanding them well in the setting of GLM seems to be a
prerequisite to understand more involved, e.g. deep learning, methods.

Statistical physics approach can be used to obtain
specific results on the high-dimensional GLM by considering data to be random independent identically distributed (iid)
matrix and modelling the labels as being created in the teacher-student
setting. The teacher generates a ground-truth vector
of weights $w$ so that $w_j \sim P_w$, $j=1,\dots,p$. The teacher then
uses this vector and data matrix $X$ to produce labels $y$ taken from  $P_{\rm
  out}(y_i| X_i w^*)$. The students then knows $X$, $y$, $P_w$
and $P_{\rm out}$ and is supposed to learn the rule the teacher uses,
i.e. ideally to learn the $w^*$. Already this setting with random input data provides interesting
insights into the algorithmic tractability of the problem as the
number of samples changes.

This line of work was pioneered by Elisabeth Gardner \cite{gardner1989three}
and actively studied in physics in the past for special cases of
$P_{\rm  out}$ and $P_W$, see e.g.~\cite{gyorgyitishby1990learning, sompolinsky1990learning,seung1992statistical}. The replica method can be used to
compute the mutual information between $X$ and $y$ in this
teacher-student model, which is related to the free energy in
physics. One can then deduce the optimal estimation error of the
vector $w^*$, as well as the optimal generalization error. A
remarkable recent progress was made in \cite{barbier2017phase} where
it has been proven that the replica method yields the correct results for the GLM
with random inputs for generic $P_{\rm  out}$ and $P_W$.
Combining these results with the analysis of the approximate
message passing algorithms \cite{javanmard2013state}, one can deduce cases where the AMP
algorithm is able to reach the optimal performance and regions where
it is not. The AMP algorithm is conjectured to
be the best of all polynomial algorithm for this case.
The teacher-student model could
thus be used by practitioners to understand how far from optimality
are general purpose
algorithms in cases where only very limited number of samples is
available.

\subsubsection{Physics results on multi-layer neural networks}

Statistical physics analysis of learning and generalization properties in deep neural networks is a challenging
task. Progress had been made in several complementary directions.

One of the influential directions involved studies of linear deep neural
networks. While linear neural networks do not have the expressive
power to represent generic functions, the learning dynamics of the
gradient descent algorithm bears
strong resemblance with the learning dynamics on non-linear
networks. At the same time the dynamics of learning in deep linear
neural networks can be described via a closed form solution
\cite{saxe2013exact}.
The learning dynamics of linear neural networks is also able to
reproduce a range of facts about generalization and over-fitting as
observed numerically in non-linear networks, see e.g. \cite{advani2017high}.

Another special case that has been analyzed in great detail is called the committee
machine, for a review see e.g. \cite{engel2001statistical}. Committee machine is
a fully-connected neural network learning a teacher-rule on random input
data with only the first layer of weights being learned, while
the subsequent ones are fixed. The theory is restricted to the limit
where the number of hidden neurons $k=O(1)$, while the
dimensionality of the input $p$ and the number of samples $n$ are both
diverge, with $n/p=\alpha=O(1)$. Both the stochastic gradient descent
(aka online) learning \cite{saad1995line,saad1995exact} and the optimal batch-learning
generalization error can be analyzed in closed form in this case
\cite{schwarze1993learning}. Recently the replica analysis of the
optimal generalization properties has been established rigorously \cite{aubin2018committee}.
A key feature of the committee machine is that it displays the
so-called {\it specialization} phase transition. When the number of
samples is small, the optimal error is achieved by a
weight-configuration that is the same for every hidden unit,
effectively implementing simple regression. Only when the number of
hidden units exceeds the {\it specialization threshold} the different
hidden units learn different weights resulting in improvement of the
generalization error. Another interesting observation about the
committee machine is that the hard phase where good generalization is
achievable information-theoretically but not tractably gets larger as
the number of hidden units grows. Committee machine was also used to analyzed the consequences of over-parametrization in neural networks in \cite{goldt2019generalisation,goldt2019dynamics}.

Another remarkable limit of two-layer neural networks was analysed in
a recent series of works \cite{mei2018mean,rotskoff2018parameters}. In
these works the networks are analysed in the limit where the number of hidden units is large,
while the dimensionality of the input is kept fixed. In this limit the
weights interact only weakly -- leading to the term {\it mean field}
-- and their evolution can be tracked via an ordinary differential
equation analogous to those studied in glassy systems \cite{dean1996langevin}.
A related, but different, treatment of the limit when the hidden layers are large is based on linearization of the dynamics around the initial condition leading to relation with Gaussian processes and kernel methods, see e.g. \cite{lee2017deep,jacot2018neural}

\subsubsection{Information Bottleneck}


Information bottleneck \cite{tishby2000information} is another concept stemming in statistical
physics that has been influential in the quest for understanding the theory behind the success of deep
learning. The theory of the information bottleneck for deep learning
\cite{tishby2015deep,shwartz2017opening} aims to quantify the
notion that layers in a neural networks are trading off between
keeping enough information about the input so that the output labels can be predicted, while forgetting as much of the unnecessary information as
possible in order to keep the learned representation concise.

 One of the interesting consequences of this information theoretic  analysis is that the traditional capacity, or expressivity dimension of the network, such as the VC dimension, is replaced
by the {\em exponent} of the mutual information between the input and the compressed hidden layer representation. This implies that every bit of representation compression is equivalent to doubling the training data in its impact on the generalization error.

The analysis of \cite{shwartz2017opening} also suggests that such
representation compression is achieved by Stochastic Gradient Descent (SGD) through diffusion in the irrelevant
dimensions of the problem. According to this, compression is achieved with {\em any units nonlinearity}
by reducing the SNR of the irrelevant dimensions, layer by layer, through the diffusion of
the weights.  An intriguing prediction of this insight is that the time to converge to good generalization scales like a negative power-law of the number of layers. The theory
also predicts a connection between the hidden layers and the bifurcations, or phase transitions, of the Information Bottleneck
representations.

While the mutual
information of the internal representations is intrinsically hard to compute directly in large neural networks,
none of the above predictions depend on explicit estimation of mutual information values.

A related line of work in
statistical physics aims to provide reliable scalable
approximations and models where the mutual information is
tractable. The mutual information can be computed exactly in linear
networks \cite{saxe2018information}. It can be reliably approximated in models of neural
networks where after learning the matrices of weights are close enough
to rotationally invariant, this is then exploited within the replica
theory in order to compute the desired mutual information \cite{gabrie2018entropy}.

\subsubsection{Landscapes and glassiness of deep learning}
\label{sec:landcapes}

Training a deep neural network is usually done via stochastic gradient
descent (SGD) in the non-convex landscape of a loss
function. Statistical physics has long experience in studies of complex energy
landscapes and and their relation to dynamical behaviour. Gradient
descent algorithms are closely related to the Langevin dynamics that
is often considered in physics. Some physics-inspired works
\cite{choromanska2015loss} became popular but were somewhat naive in
exploring this analogy.

Interesting insight on the relation between glassy dynamics and
learning in deep neural networks is presented in
\cite{baity2018comparing}. In particular the role of
over-parameterization in making the landscape look less glassy is
highlighted and contrasted with the under-parametrized networks.

Another intriguing line of work that relates learning in neural networks
to properties of landscapes is explored in \cite{baldassi2015subdominant,baldassi2016unreasonable}.
This work is based on realization that in the simple model of binary perceptron
learning dynamics ends in a part of the weight-space that has many
low-loss close-by configurations. It goes on to suggest that learning favours these wide parts in the
space of weights, and argues that this might explain why
algorithms are attracted to wide local minima and why by doing so
their generalization properties improve. An interesting spin-off of this theory is a variant of the
stochastic gradient descent algorithm suggested in
\cite{chaudhari2016entropy}.

\subsection{Applications of ML in Statistical Physics}
\label{subsec:ml_in_stat_phys}

When a researcher in theoretical physics encounters deep neural
networks where the early layers are learning to represent the input
data at a finer scale than the later layers, she immediately thinks
about renormalization group as used in physics in order to extract
macroscopic behaviour from microscopic rules. This analogy was
explored for instance in \cite{beny2013deep,mehta2014exact}.
Analogies between renormalization group and the principle component
analysis were reported in \cite{bradde2017pca}.

A natural idea is to use neural networks in order to learn new renormalization
schemes. First attempts in this direction appeared in
\cite{koch2018mutual,li2018neural}. However, it remains to be shown whether this can
lead to new physical discoveries in models that were not well understood previously.

Phase transitions are boundaries between different phases of
matter. They are usually determined using order parameters. In some systems it
is not a priori clear how to determine the proper order parameter. A
natural idea is that a neural networks may be able to learn appropriate
order parameters and locate the phase transition without a priori
physical knowledge. This idea was explored in
\cite{carrasquilla2017machine, van2017learning, tanaka2017detection, morningstar2017deep} in a range of models using configurations sampled uniformly from the model of interest
(obtained using Monte Carlo simulations) in different phases or at different temperatures
and using supervised learning in order to classify the configurations to
their phases. Extrapolating to configurations not used in the training
set plausibly leads to determination of the phase transitions in the
studied models.
These general guiding principles have been used in a large number of applications to analyze both synthetic and experimental data. Specific cases in the context of many-body quantum physics are detailed in Section \ref{subsec-class_manybody}.

Detailed understanding of the limitations of these
methods in terms of identifying previously unknown order parameters, as well as understanding whether they can reliably
distinguish between a true thermodynamic phase transitions and a mere cross-over are yet to be clarified. Experiments presented on the
Ising model in \cite{mehta2018high}
provide some preliminary thoughts in that direction. Some underlying mechanisms are discussed in \cite{kashiwa2019phase}. Kernel based learning method for learning phases in frustrated magnetic materials that is more easily interpretable and able to identify complex order parameters is introduced and studied in
\cite{greitemann2019probing,liu2019learning}.

Disordered and
glassy solids where identifications of the order parameter is
particularly challenging were also studied. In particular \cite{ronhovde2011detecting,nussinov2016inference} use multi-scale network clustering methods to identify spatial and spatio-temporal structures in glasses,
\cite{cubuk2015identifying} learn to identify structural flow defects, and \cite{schoenholz2017relationship} argues to identify a parameter that captures the history dependence of the disordered system.


In an ongoing effort to go beyond the limitations of supervised learning to classify phases and identify phase transitions, several direction towards unsupervised learning are begin explored. For instance, in \cite{wetzel2017unsupervised} for
the Ising and XY model, in \cite{wang2017machine,wang2018machine}
for frustrated spin systems.
The work of \cite{martiniani2017quantifying} explores the direction of identifying
phases from simple compression of the underlying configurations.

Machine learning also provides exciting set of tools to study, predict and control non-linear dynamical systems. For instance  \cite{pathak2017using,pathak2018model} used  recurrent neural networks called an echo state networks or reservoir computers \cite{jaeger2004harnessing} to predict the trajectories of a chaotic dynamical system and of models used for weather prediction.
The authors of \cite{reddy2016learning,reddy2018glider} used reinforcement learning to teach an autonomous glider to literally soar like a bird, using thermals in the atmosphere.

\subsection{Outlook and Challenges}

The described methods of statistical physics are quite powerful in dealing with high-dimensional data sets and models. The largest difference between traditional learning theories and the theories coming from statistical physics is that the later are often based on toy generative models of data. This leads to solvable models in the sense that quantities of interest such as achievable errors can be computed in a closed form, including constant terms. This is in contrast with aims in the mainstream learning theory that aims to provide worst case bounds on error under general assumptions on the setting (data structure, or architecture). These two approaches are complementary and ideally will meet in the future once we understand what are the key conditions under which practical cases are close to worse cases, and what are the right models of realistic data and functions.

The next challenge for the statistical physics approach is to formulate and solve models that are in some kind of {\it universality class} of the real settings of interest. Meaning that they reproduce all important aspects of the  behaviour that is observed in practical application of neural networks. For this we need to model the input data no longer as iid vectors, but for instance as outputs from a generative neural network as in \cite{gabrie2018entropy}, or as perceptual manifolds as in \cite{chung2018classification}. The teacher network that is producing the labels (in an supervised setting) needs to model suitably the correlation between the structure in the data and the label. We need to find out how to analyze the (stochastic) gradient descent algorithm and its relevant variants. Promising works in this direction, that rely of the dynamic mean-field theory of glasses are \cite{mannelli2018marvels,mannelli2019passed}. We need to generalize the existing methodology to multi-layer networks with extensive width of hidden layers.

Going back to the direction of using machine learning for physics, the full potential of ML in research of non-linear dynamical systems and statistical physics is yet to be uncovered. The above mentioned works certainly provide an exciting appetizer.


\section{Particle Physics and Cosmology}
\label{sec:particle}

A diverse portfolio of on-going and planned experiments is well poised to explore the universe from the unimaginably small world of fundamental particles to the awe inspiring scale of the universe.
Experiments like the Large Hadron Collider (LHC) and the Large Synoptic Survey Telescope (LSST) deliver enormous amounts of data to be compared to the predictions of specific theoretical models. Both areas have well established physical models that serve as null hypotheses: the standard model of particle physics and $\Lambda$CDM cosmology, which includes cold dark matter and a cosmological constant $\Lambda$. Interestingly, most alternate hypotheses considered are formulated in the same theoretical frameworks, namely quantum field theory and general relativity. Despite such sharp theoretical tools, the challenge is still daunting as the expected deviations from the null are expected to be incredibly tiny and revealing such subtle effects requires a robust treatment of complex experimental apparatuses. Complicating the statistical inference is that the most high-fidelity predictions for the data do not come in the from simple closed-form equations, but instead in complex computer simulations.

Machine learning is making waves in particle physics and cosmology as it offers a suit of techniques to confront these challenges and a new perspective that motivates bold new strategies. The excitement spans the theoretical and experimental aspects of these fields and includes both applications with immediate impact as well as the prospect of more transformational changes in the longer term.

\subsection{The role of the simulation}

An important aspect of the use of machine learning in particle physics and cosmology is the use of computer simulations to generate  samples of labeled training data $\{X_\mu, y_\mu\}_{\mu=1}^n$.
For example, when the target $y$ refers to a particle type, particular scattering process, or parameter appearing in the fundamental theory, it can often be specified directly in the simulation code so that the simulation directly samples $X \sim p(\cdot | y)$. In other cases, the simulation is not directly conditioned on $y$, but provides samples $(X,Z) \sim p(\cdot)$, where $Z$ are latent variables that describe what happened inside the simulation, but which are not observable in an actual experiment. If the target label can be computed from these latent variables via a function $y(Z)$, then  labeled training data $\{X_\mu, y(Z_\mu)\}_{\mu=1}^n$ can also be created from the simulation.  The use of high-fidelity simulations to generate labeled training data has not only been the key to early successes of supervised learning in these areas, but also the focus of research addressing the shortcomings of this approach.

Particle physicists have developed a suite of high-fidelity simulations that are hierarchically composed to describe interactions across a huge range of length scales. The components of these simulations include Feynman diagrammatic perturbative expansion of quantum field theory, phenomenological models for complex patterns of radiation, and detailed models for interaction of particles with matter in the detector. While the resulting simulation has high fidelity, the simulation itself has free parameters to be tuned and number of residual uncertainties in the simulation must be taken into account in down-stream analysis tasks.

Similarly, cosmologists can simulate the evolution of the universe at different length scales using general relativity and relevant non-gravitational effects of matter and radiation that becomes increasingly important during structure formation. There is a rich array of approximations that can be made in specific settings that provide enormous speedups compared to the computationally expensive $N$-body simulations of billions of massive objects that interact gravitationally, which become prohibitively expensive once non-gravitational feedback effects are included.

Cosmological simulations generally involve deterministic evolution of stochastic initial conditions due to primordial quantum fluctuations. The $N$-body simulations are very expensive, so there are relatively few simulations, but they cover a large space-time volume that is statistically isotropic and homogeneous at large scales. In contrast, particle physics simulations are stochastic throughout from the initial high-energy scattering to the low-energy interactions in the detector. Simulations for high-energy collider experiments can run on commodity hardware in a parallel manner, but the physics goals requires enormous numbers of simulated collisions.

Because of the critical role of the simulation in these fields, much of the recent research in machine learning is related to simulation in one way or another. These goals of these recent works are to:
\begin{itemize}
   \item develop techniques that are more data efficient by incorporating domain knowledge directly into the machine learning models;
   \item incorporate the uncertainties in the simulation into the training procedure;
   \item develop weakly supervised procedures that can be applied to real data and do not rely on the simulation;
   \item develop anomaly detection algorithms to find anomalous features in the data without simulation of a specific signal hypothesis;
   \item improve the tuning of the simulation, reweight or adjust the simulated data to better match the real data, or use machine learning to model residuals between the simulation and the real data;
   \item learn fast neural network surrogates for the simulation that can be used to quickly generate synthetic data;
   \item develop approximate inference techniques that make efficiently use of the simulation; and
   \item learn fast neural network surrogates that can be used directly for statistical inference.
\end{itemize}

\subsection{Classification and regression in particle physics}
\label{sec:Class_part}

Machine learning techniques have been used for decades in experimental particle physics to aid  particle identification and event selection, which can be seen as classification tasks. Machine learning has also been used for reconstruction, which can be seen as a regression task.
Supervised learning is used to train a predictive model based on large number of labeled training samples $\{X_\mu, y_\mu\}_{\mu=1}^n$, where $X$ denotes the input data and $y$ the target label.
In the case of particle identification, the input features~$X$ characterize localized energy deposits in the detector and the label $y$ refers to one of a few particle species (e.g. electron, photon, pion, etc.). In the reconstruction task, the same type of sensor data $X$ are used, but the target label $y$ refers to the energy or momentum of the particle responsible for those energy deposits.
These algorithms are applied to the bulk data processing of the LHC data.

Event selection, refers to the task of selecting a small subset of the collisions that are most relevant for a targeted analysis task. For instance, in the search for the Higgs boson, supersymmetry, and dark matter data analysts must select a small subset of the LHC data that is consistent with the features of these hypothetical "signal" processes. Typically these event selection requirements are also satisfied by so-called "background" processes that mimic the features of the signal either due to experimental limitations or fundamental quantum mechanical effects. Searches in their simplest form reduce to comparing the number of events in the data that satisfy these requirements to the predictions of a background-only null hypothesis and signal-plus-background alternate hypothesis. Thus, the more effective the event selection requirements are at rejecting background processes and accept signal processes, the more powerful the resulting statistical analysis will be. Within high-energy physics, machine learning classification techniques have traditionally been referred to as \textit{multivariate analysis} to emphasize the contrast to traditional techniques based on simple thresholding (or ``cuts'') applied to carefully selected or engineered features.

In the 1990s and early 2000s simple feed-forward neural networks were commonly used for these tasks. Neural networks were largely displaced by Boosted Decision Trees (BDTs) as the go-to for classification and regression tasks for more than a decade~\cite{Breiman1984ClassificationAR, Freund:1997xna, Roe:2004na}. Starting around 2014, techniques based on deep learning emerged and were demonstrated to be significantly more powerful in several applications (for a recent review of the history, see Refs.~\cite{Guest:2018yhq, Radovic:2018dip}).

Deep learning was first used for an event-selection task targeting hypothesized particles from theories beyond the standard model. It not only out-performed boosted decision trees, but also did not require engineered features to achieve this impressive performance~\cite{Baldi:2014kfa}. In this proof-of-concept work, the network was a deep multi-layer perceptron trained with a very large training set using a simplified detector setup.  Shortly after, the idea of a parametrized classifier was introduced in which the concept of a binary classifier was extended to a situation where the $y=1$ signal hypothesis is lifted to a composite hypothesis that is parameterized continuously, for instance, in terms of the mass of a hypothesized particle~\cite{Baldi:2016fzo}.

\subsubsection{Jet Physics}
The most copious interactions at hadron colliders such as the LHC produce high energy quarks and gluons in the final state. These quarks and gluons radiate more quarks and gluons that eventually combine into color-neutral composite particles due to the phenomena of confinement. The resulting collimated  spray of mesons and baryons that strike the detector is collectively referred to as a jet. Developing a useful characterization of the structure of a jet that are theoretically robust and that can be used to test the predictions of quantum chromodynamics (QCD) has been an active area of particle physics research for decades. Furthermore, many scenarios for physics Beyond the Standard Model predict the production of particles that decay into two or more jets. If those unstable particles are produced with a large momentum, then the resulting jets are boosted such that the jets overlap into a single fat jet with nontrivial substructure. Classifying these boosted or fat jets from the much more copiously produced jets from standard model processes involving quarks and gluons is an area that can significantly improve the physics reach of the LHC. More generally, identifying the progenitor for a jet is a classification task that is often referred to as jet tagging.

Shortly after the first applications of deep learning for event selection, deep convolutional networks were used for the purpose of jet tagging, where the low-level detector data lends itself to an image-like representation~\cite{deOliveira:2015xxd, baldi2016jet}. While machine learning techniques have been used within particle physics for decades, the practice has always been restricted to input features $X$ with a fixed dimensionality. One challenge in jet physics is that the natural representation of the data is in terms of particles, and the number of particles associated to a jet varies. The first application of a recurrent neural network in particle physics was in the context of flavor tagging~\cite{Guest:2016iqz}.
More recently, there has been an explosion of research into the use of different network architectures including recurrent networks operating on sequences, trees, and graphs (see Ref.~\cite{Larkoski:2017jix} for a recent review for jet physics). This includes hybrid approaches that leverage domain knowledge in the design of the architecture. For example, motivated by techniques in natural language processing, recursive networks were designed that operate over tree-structures created from a class of jet clustering algorithms~\cite{Louppe:2017ipp}.  Similarly, networks have been developed motivated by invariance to permutations on the particles presented to the network and stability to details of the radiation pattern of particles,
~\cite{Komiske:2017aww, Komiske:2018cqr}. Recently, comparisons of the different approaches for specific benchmark problems have been organized~\cite{Kasieczka:2019dbj}.

In addition to classification and regression, machine learning techniques have been used for density estimation and modeling smooth spectra where an analytical form is not well motivated and the simulation has significant uncertainties~\cite{Frate:2017mai}. The work also allows one to model alternative signal hypotheses with a diffuse prior instead of a specific concrete physical model. More abstractly, the Gaussian process in this work is being used to model the intensity of inhomogeneous Poisson point process, which is a scenario that is found in particle physics, astrophysics, and cosmology. One interesting aspect of this line of work is that the Gaussian process kernels can be constructed using compositional rules that correspond clearly to the causal model physicists intuitively use to describe the observation, which aids in interpretability~\cite{duvenaud2013structure}.

\subsubsection{Neutrino physics}

Neutrinos interact very feebly with matter, thus the experiments require large detector volumes to achieve appreciable interaction rates. Different types of interactions, whether they come from different species of neutrinos or background cosmic ray processes, leave different patterns of localized energy deposits in the detector volume.  The detector volume is homogeneous, which motivates the use of convolutional neural networks.

The first application of a deep convolutional network in the analysis of data from a particle physics experiment was in the context of the NO$\nu$A experiment, which uses scintillating mineral oil. Interactions in NO$\nu$A lead to the production of light, which is imaged from two different vantage points. NO$\nu$A developed a convolutional network that simultaneously processed these two images~\cite{Aurisano:2016jvx}. Their network improves the efficiency (true positive rate) of selecting electron neutrinos by 40\% for the same purity. This network has been used in searches for the appearance of electron neutrinos and for the hypothetical sterile neutrino.

Similarly, the MicroBooNE experiment detects neutrinos created at Fermilab. It uses 170 ton liquid-argon time projection chamber. Charged particles ionize the liquid argon and the ionization electrons drift through the volume to three wire planes. The resulting data is processed and represented by a 33-megapixel image, which is dominantly populated with noise and only very sparsely populated with legitimate energy deposits. The MicroBooNE collaboration used a FasterRCNN~\cite{ren2015faster} to identify and localize neutrino interactions with bounding boxes~\cite{Acciarri:2016ryt}. This success is important for future neutrino experiments based on liquid-argon time projection chambers, such as the Deep Underground Neutrino Experiment (DUNE).

In addition to the relatively low energy neutrinos produced at accelerator facilities, machine learning has also been used to study high-energy neutrinos with the IceCube observatory located at the south pole. In particular, 3D convolutional and graph neural networks have been applied to a signal classification problem. In the latter approach, the detector array is modeled as a graph, where vertices are sensors and edges are a learned function of the sensors' spatial coordinates. The graph neural network was found to outperform both a traditional-physics-based method as well as classical 3D convolutional neural network~\cite{2018arXiv180906166C}.


\subsubsection{Robustness to systematic uncertainties}

Experimental particle physicists are keenly aware that the simulation, while incredibly accurate, is not perfect. As a result, the community has developed a number of strategies falling roughly in two broad classes. The first involves incorporating the effect of mis-modeling when the simulation is used for training. This involves either propagating the underlying sources of uncertainty (e. g.  calibrations, detector response, the quark and gluon composition of the proton, and the impact of higher-order corrections from perturbation theory, etc.)  through the simulation and analysis chain. For each of these sources of uncertainty, a nuisance parameter $\nu$ is included, and the resulting statistical model $p(X | y, \nu)$ is parameterized by these nuisance parameters. In addition, the likelihood function for the data is augmented with a term  $p(\nu)$ representing the uncertainty in these sources of uncertainty, as in the case of a penalized maximum likelihood analysis.
In the context of machine learning, classifiers and regressors are typically trained using data generated from a nominal simulation $\nu=\nu_0$, yielding a predictive model $f(X | \nu_0)$. Treating this predictive model as fixed, it is possible to propagate the uncertainty in $\nu$ through $f(X | \nu_0)$ using the model $p(X | y, \nu)p(\nu)$. However, the down-stream statistical analysis based on this approach is not optimal since the predictive model was not trained taking into account the uncertainty on $\nu$.

In machine learning literature, this situation is often referred to as {\it covariate shift} between two domains represented by the training distribution $\nu_0$ and the target distribution $\nu$. Various techniques for domain adaptation exist to train classifiers that are robust to this change, but they tend to be restricted to binary domains $\nu \in \{\textrm{train}, \textrm{target} \}$. To address this problem, an adversarial training technique was developed that extends domain adaptation to domains parametrized by $\nu \in \mathbb{R}^q$~\cite{Louppe:2016ylz}. The adversarial approach encourages the network to learn a \textit{pivotal} quantity, where $p(f(X)|y, \nu)$ is independent of $\nu$, or equivalently $p(f(X), \nu | y) = p(f(X) | y) p(\nu)$. This adversarial approach has also been used in the context of algorithmic fairness, where one desires to train a classifiers or regressor that is independent of (or decorrelated with) specific continuous attributes or observable quantities. For instance, in jet physics one often would like a jet tagger that is independent of the jet invariant mass~\cite{Shimmin:2017mfk}. Previously, a different algorithm called \texttt{uboost} was developed  to achieve similar goals for boosted decision trees~\cite{Stevens:2013dya,Rogozhnikov:2014zea}.

The second general strategy used within particle physics to cope with systematic mis-modeling in the simulation is to avoid using the simulation for modeling the distribution $p(X|y)$.  In what follows, let $R$ denote an index over various subsets of the data satisfying corresponding selection requirements. Various data-driven strategies have been developed to relate distributions of the data in control regions, $p(X|y, R=0)$, to distributions in regions of interest, $p(X|y, R=1)$. These relationships also involve the simulation, but the art of this approach is to base those relationships on aspects of the simulation that are considered robust. The simplest example is estimating the distribution $p(X|y,R=1)$ for a specific process $y$ by identifying a subset of the data $R=0$ that is dominated by $y$ and $p(y|R=0) \approx 1$. This is an extreme situation that is limited in applicability.

Recently, weakly supervised techniques have been developed that only involve identifying regions where only the class proportions are known or assuming that the relative probabilities $p(y|R)$ are not linearly dependent~\cite{Komiske:2018oaa,Metodiev:2017vrx} . The techniques also assume that the distributions $p(X|y,R)$ are independent of $R$, which is reasonable in some contexts and questionable in others. The approach has been used to train jet taggers that discriminate between quarks and gluons, which is an area where the fidelity of the simulation is no longer adequate and the assumptions for this method are reasonable. This weakly-supervised, data-driven approach is a major development for machine learning for particle physics, though it is limited to a subset of problems. For example, this approach is not applicable if one of the target categories $y$ corresponds to a hypothetical particle that may not exist or be present in the data.

%
%


\subsubsection{Triggering}

Enormous amounts of data must be collected by collider experiments such as the LHC, because the phenomena being targeted are exceedingly rare. The bulk of the collisions involve phenomena that have previously been studied and characterized, and the data volume associated with the full data stream is impractically large. As a result, collider experiments use a real-time data-reduction system referred to as a trigger. The trigger makes the critical decision of which events to keep for future analysis and which events to discard. The ATLAS and CMS experiments retain only about 1 out of every 100,000 events. Machine learning techniques are used to various degrees in these systems. Essentially, the same particle identification (classification) tasks appears in this context, though the computational demands and performance in terms of false positives and negatives are different in the real-time environment.

The LHCb experiment has been a leader in using machine learning techniques in the trigger. Roughly 70\% of the data selected by the LHC trigger is selected by machine learning algorithms. Initially, the experiment used a boosted decision tree for this purpose~\cite{Gligorov:2012qt}, which was later replaced by the MatrixNet algorithm developed by Yandex~\cite{Likhomanenko:2015aba}.

The Trigger systems often use specialized hardware and firmware, such as   field-programmable gate arrays (FPGAs). Recently, tools have been developed to streamline the compilation of machine learning models for FPGAs to target the requirements of these real-time triggering systems~\cite{Duarte:2018ite,Tsaris:2018myg}.

\subsubsection{Theoretical particle physics}

While the bulk of machine learning in particle physics and cosmology are focused on analysis of observational data, there are also examples of using machine learning as a tool in theoretical physics. For instance, machine learning has been used to characterize the  landscape of string theories~\cite{Carifio:2017bov}, to identify the phase transitions of quantum chromodynamics (QCD)~\cite{Pang:2016vdc}, and to study the the AdS/CFT correspondence~\cite{PhysRevD.98.106014,Hashimoto:2018ftp}.
Some of this work is more closely connected to the use of machine learning as a tool in condensed matter or many-body quantum physics.  Specifically, deep learning has been used in the context of lattice QCD (LQCD). In an exploratory work in this direction, deep neural networks were used to predict the parameters in the QCD Lagrangian from lattice configurations~\cite{Shanahan:2018vcv}. This is needed for a number of multi-scale action-matching approaches, which aim to improve the efficiency of the computationally intensive LQCD calculations. This problem was setup as a regression task, and one of the challenges is that there are relatively few training examples. Additionally, machine learning techniques are being used to reduce the autocorrelation time in the Markov Chains~\cite{Tanaka:2017niz,Albergo:2019eim} In order to solve this task with few training examples it is important to leverage the known space-time and local gauge symmetries in  the lattice data. Data augmentation is not a scalable solution given the richness of the symmetries. Instead the authors performed feature engineering that imposed gauge symmetry and space-time translational invariance. While this approach proved effective, it would be desirable to consider a richer class of networks, that are equivariant (or covariant) to the symmetries in the data (such approaches are discussed in Sec.~\ref{S:hep-outlook}). A continuation of this work is being supported by the Argon Leadership Computing Facility. The a new Intel-Cray system Aurora, will be capable of over 1 exaflops and specifically is aiming at problems that combine traditional high performance computing with modern machine learning techniques.

\subsection{Classification and regression in cosmology}

\subsubsection{Photometric Redshift}

Due to the expansion of the universe the distant luminous objects are redshifted, and the distance-redshift relation is a fundamental component of observational cosmology. Very precise redshift estimates can be obtained through spectroscopy; however, such spectroscopic surveys are expensive and time consuming. Photometric surveys based on broadband photometry or imaging in a few color bands give a coarse approximation to the spectral energy distribution. Photometric redshift refers to the regression task of estimating redshifts from photometric data. In this case, the ground truth training data comes from precise spectroscopic surveys.


The traditional approaches to photometric redshift is based on template fitting methods~\cite{2000ApJ...536..571B,2008ApJ...686.1503B,2006MNRAS.372..565F}. For more than a decade cosmologists have also used machine learning methods based on neural networks and boosted decision trees for photometric redshift~\cite{2003MNRAS.339.1195F,2004PASP..116..345C,2013MNRAS.432.1483C}. One interesting aspect of this body of work is the effort that has been placed to go beyond a point estimate for the redshift. Various approaches exist to determine the uncertainty on the redshift estimate and to obtain a posterior distribution.

While the training data are not generated from a simulation, there is still a concern that the distribution of the training data may not be  representative of the distribution of data that the models will be applied to. This type of covariate shift results from various selection effects in the spectroscopic survey and subtleties in the photometric surveys. The Dark Energy Survey considered a number of these approaches and established a validation process to evaluate them critically~\cite{Bonnett:2015pww}. Recently there has been work to use hierarchical models to build in additional causal structure in the models to be robust to these differences. In the language of machine learning, these new models  aid in transfer learning and domain adaptation. The hierarchical models also aim to combine the interpretability of traditional template fitting approaches and the flexibility of the machine learning models~\cite{2018arXiv180701391L}.

\subsubsection{Gravitational lens finding and parameter estimation}

One of the most striking effects of general relativity is gravitatioanl lensing, in which a massive foreground object warps the image of a background object. Strong gravitational lensing occurs, for example, when a massive foreground galaxy is nearly coincident on the sky with a background source. These events are a powerful probe of the dark matter distribution of massive galaxies and can provide valuable cosmological constraints. However, these systems are rare, thus a scalable and reliable lens finding system is essential to cope with large surveys such as LSST, Euclid, and WFIRST. Simple feedfoward, convolutional and residual neural networks (ResNets) have been applied to this supervised classification problem~\cite{2007ApJ...660.1176E,marshall2009automated,2018MNRAS.473.3895L}. In this setting, the training data came from simulation using PICS (Pipeline for Images of Cosmological Strong) lensing~\cite{Li_2016} for the strong lensing ray-tracing and LensPop~\cite{Collett_2015} for mock LSST observing. Once identified, characterizing the lensing object through maximum likelihood estimation is a computationally intensive non-linear optimization task. Recently, convolutional networks have been used to quickly estimate the parameters of the Singular Isothermal Ellipsoid density profile, commonly used to model strong lensing systems~\cite{Hezaveh:2017sht}.

\subsubsection{Other examples}

In addition to the examples above, in which the ground truth for an object is relatively unambiguous with a more labor-intensive approach, cosmologists are also leveraging machine learning to infer quantities that involve unobservable latent processes or the parameters of the fundamental cosmological model.

\begin{figure*}
    \centering
\includegraphics[width=.45\textwidth]{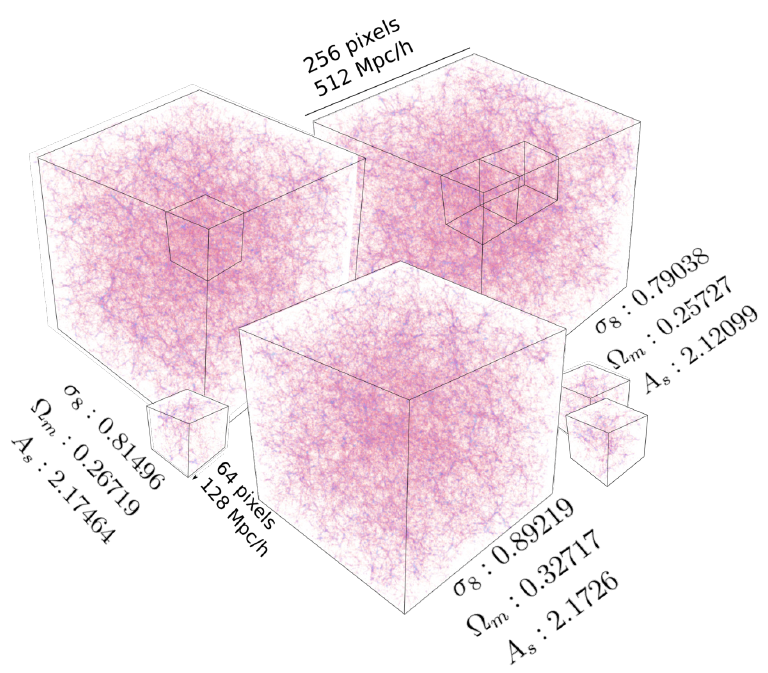}
    \caption{Dark matter distribution in three cubes produced using different sets of parameters. Each cube is divided into small sub- cubes for training and prediction. Note that although cubes in this figure are produced using very different cosmological parameters in our constrained sampled set, the effect is not visually discernible.  Reproduced from ~\cite{2017arXiv171102033R}.}
    \label{fig:dm-sim}
\end{figure*}

For example, 3D convolutional networks have been trained to predict fundamental cosmological parameters based on the dark matter spatial distribution~\cite{2017arXiv171102033R} (see Fig.~\ref{fig:dm-sim}). In this proof-of-concept work, the networks were trained using computationally intensive $N$-body simulations for the evolution of dark matter in the universe assuming specific values for the 10 parameters in the standard $\Lambda$CDM cosmology model. In real applications of this technique to visible matter, one would need to model the bias and variance of the visible tracers with respect to the underlying dark matter distribution. In order to close this gap, convolutional networks have been trained to learn a fast mapping between the dark matter and visible galaxies~\cite{2019arXiv190205965Z}, allowing for a trade-off between simulation accuracy and computational cost. One challenge of this work, which is common to applications in solid state physics, lattice field theory, and many body quantum systems, is that the simulations are computationally expensive and thus there are relatively few statistically independent realizations of the large simulations $X_\mu$. As deep learning tends to require large labeled training data-sets, various types of subsampling and data augmentation approaches have been explored to ameliorate the situation. An alternative approach to subsampling is the so-called Backdrop, which provides stochastic gradients of the loss function even on individual samples by introducing a stochastic masking in the backpropagation pipeline~\cite{Golkar:2018xkw}.

Inference on the fundamental cosmological model also appears in a classification setting. In particular, modified gravity models with massive neutrinos can mimic the predictions for weak-lensing observables predicted by the standard $\Lambda$CDM  model. The degeneracies that exist when restricting the $X_\mu$ to second-order statistics can be broken by incorporating higher-order statistics or other rich representations of the weak lensing signal. In particular, the authors of ~\cite{2018arXiv181011030P} constructed a novel representation of the wavelet decomposition of the weak lensing signal as input to a convolutional network. The resulting approach was able to discriminate between previously degenerate models with 83\%–100\% accuracy.

Deep learning has also been used to estimate the mass of galaxy clusters, which are the largest gravitationally bound structures in the universe and a powerful cosmological probe. Much of the mass of these galaxy clusters comes in the form of dark matter, which is not directly observable. Galaxy cluster masses can be estimated via gravitational lensing, X-ray observations of the intra-cluster medium, or through dynamical analysis of the cluster's galaxies.  The first use of machine learning for a dynamical cluster mass estimate was performed using Support Distribution Machines~\cite{DBLP:journals/corr/abs-1202-0302} on a dark-matter-only simulation~\cite{2015ApJ...803...50N,2016ApJ...831..135N}.
A number of non-neural network algorithms including Gaussian process regression (kernel ridge regression), support vector machines, gradient boosted tree regressors, and others have been applied to this problem using the MACSIS simulations~\cite{10.1093/mnras/stw2722} for training data. This simulation goes beyond the dark-matter-only simulations and incorporates the impact of various astrophysical processes and allows for the development of a realistic processing pipeline that can be applied to observational data. The need for an accurate, automated mass estimation pipeline is motivated by large surveys such as eBOSS, DESI, eROSITA, SPT-3G, ActPol, and Euclid. The authors found that compared to the traditional $\sigma-M$ relation the predicted-to-true mass ratio using machine learning techniques is reduced by a factor of 4~\cite{2019MNRAS.484.1526A}. Most recently, convolutional neural networks have been used to mitigate systematics in the virial scaling relation, further improving dynamical mass estimates~\cite{Ho:2019zap}.
Convolutional neural networks have also been used to estimate cluster masses with synthetic (mock) X-ray observations of galaxy clusters, where the authors find the scatter in the predicted mass is reduced compared to traditional X-ray luminosity based methods~\cite{2018arXiv181007703N}.



\subsection{Inverse Problems and Likelihood-free inference}
\label{sec:likelihood_free}

As stressed repeatedly, both particle physics and cosmology are characterized by well motivated, high-fidelity forward simulations. These forward simulations are either intrinsically stochastic -- as in the case of the probabilistic decays and interactions found in particles physics simulations -- or they are deterministic -- as in the case of gravitational lensing or N-body gravitational simulations. However, even deterministic physics simulators usually are followed by a probabilistic description of the observation based on Poisson counts or a model for instrumental noise. In both cases, one can consider the simulation as implicitly defining the distribution $p(X, Z | y)$, where $X$ refers to the observed data, $Z$ are unobserved latent variables that take on random values inside the simulation, and $y$ are parameters of the forward model such as coefficients in a Lagrangian or the 10 parameters of $\Lambda$CDM cosmology. Many scientific tasks can be characterized as inverse problems where one wishes to infer $Z$ or $y$ from $X=x$. The simplest cases that we have considered are classification where $y$ takes on categorical values and regression where $y \in \mathbb{R}^d$. The point estimates $\hat{y}(X=x)$ and $\hat{Z}(X=x)$ are useful, but in scientific applications we often require uncertainty on the estimate.


In many cases, the solution to the inverse problem is ill-posed, in the sense that small changes in $X$ lead to large changes in the estimate. This implies the estimator will have high variance.  In some cases the forward model is equivalent to a linear operator and the maximum likelihood estimate $\hat{y}_\textrm{MLE}{(X)}$ or $\hat{Z}_\textrm{MLE}(X)$ can be expressed as a matrix inversion. In that case, the instability of the inverse is related to the matrix for the forward model being  poorly conditioned. While the maximum likelihood estimate may be unbiased, it tends to be high variance. Penalized maximum likelihood, ridge regression (Tikhonov regularization), and Gaussian process regression are closely related approaches to the bias-variance trade-off.

Within particle physics, this type of problem is often referred to as \textit{unfolding}. In that case, one is often interested in the distribution of some kinematic property of the collisions prior to the detector effects, and $X$ represents a smeared version of this quantity after folding in the detector effects. Similarly, estimating the parton density functions that describe quarks and gluons inside the proton can be cast as an inverse problem of this sort~\cite{Forte:2002fg,Ball:2014uwa}. Recently, both neural networks and Gaussian processes with more sophisticated, physically inspired kernels have been applied to these problems~\cite{Frate:2017mai,Bozson:2018asz}. In the context of cosmology, an example inverse problem is to denoise the Laser Interferometer Gravitational-Wave Observatory (LIGO) time series to the underlying waveform from a gravitational wave \cite{Shen:2019ohi}. Generative Adversarial Networks (GANs) have even been used in the context of inverse problems where they were used to denoise and recover images of galaxies beyond naive deconvolution limits~\cite{schawinski_generative_2017}. Another example involves estimating the image of a background object prior to being gravitationally lensed by a foreground object. In this case, describing a physically motivated prior for the background object is difficult. Recently, recurrent inference machines~\cite{putzky2017recurrent} have been introduced as way to implicitly learn a prior for such inverse problems, and they have successfully been applied to strong gravitational lensing~\cite{2018arXiv180800011M,2019arXiv190101359M}.



A more ambitious approach to inverse problems involves providing detailed probabilistic characterization of $y$  given $X$.  In the frequentist paradigm one would aim to characterize the likelihood function $L(y) = p(X=x | y)$, while in a Bayesian formalism one would wish to characterize the posterior $p(y|X=x) \propto p(X=x|y) p(y)$. The analogous situation happens for inference of latent variables $Z$ given $X$. Both particle physics and cosmology have well-developed approaches to statistical inference based on detailed modeling of the likelihood, Markov Chain Monte Carlo (MCMC)~\citep{ForemanMackey:2012ig}, Hamiltonian Monte Carlo, and variational inference~\citep{2016ascl.soft04008L,2018arXiv180300113R,Jain:2018bbr}. However, all of these approaches require that the likelihood function is tractable.

\subsubsection{Likelihood-free Inference}

Somewhat surprisingly, the probability density, or likelihood, $p(X=x|y)$ that is implicitly defined by the simulator is often intractable.
Symbolically, the probability density  can be written
$p(X|y) = \int p(X,Z | y) dZ$,  where~$Z$ are the latent variables of the simulation.
The latent space of state-of-the-art simulations is enormous and highly structured, so this integral cannot be performed analytically. In simulations of a single collision at the LHC, $Z$ may have hundreds of millions of components. In practice, the simulations are often based on Monte Carlo techniques and generate samples $(X_\mu, Z_\mu) \sim p(X,Z | y)$ from which the density can be estimated. The challenge is that if $X$ is high-dimensional it is difficult to accurately estimate those densities. For example, naive histogram-based approaches do not scale to high dimensions and kernel density estimation techniques are only trustworthy to around 5-dimensions. Adding to the challenge is that the distributions have a large dynamic range, and the interesting physics often sits in the tails of the distributions.

The intractability of the likelihood implicitly defined by the simulations is a foundational problem not only for particle physics and cosmology, but many other areas of science as well including epidemiology and phylogenetics. This has motivated the development of so-called \textit{likelihood-free inference} algorithms, which only require the ability to generate samples from the simulation in the forward mode.

One prominent technique, is Approximate Bayesian Computation (ABC). In ABC one performs Bayesian inference using MCMC or a rejection sampling approach in which the likelihood is approximated by the probability $p(\rho(X,x) < \epsilon)$, where $x$ is the observed data to be conditioned on, $\rho(x',x)$ is some distance metric between $x$ and the output of the simulator $x'$, and $\epsilon$ is a tolerance parameter. As $\epsilon  \to 0$, one recovers exact Bayesian inference; however, the efficiency of the procedure vanishes. One of the challenges for ABC, particularly for high-dimensional~$x$,  is the specification of the distance measure $\rho(x',x)$ that maintains reasonable acceptance efficiency without degrading the quality of the inference~\cite{beaumont2002approximate,marjoram2003markov,sisson2011likelihood,
sisson2007sequential,marin2012approximate}. This approach to estimating the likelihood is quite similar to the traditional practice in particle physics of using histograms or kernel density estimation to approximate $\hat{p}(x|y)\approx {p}(x|y)$. In both cases, domain knowledge is required to identify useful summary in order to reduce the dimensionality of the data.  An interesting extension of the ABC technique utilizes universal probabilistic programming. In particular, a technique known as inference compilation is a sophisticated form of importance sampling in which a neural network controls the random number generation in the probabilistic program to bias the simulation to produce outputs $x'$ closer to the observed $x$~\cite{le2017inference}.

The term ABC is often used synonymously with the more general term likelihood-free inference; however, there are a number of other approaches that involve learning an approximate likelihood or likelihood ratio that is used as a  surrogate for the intractable likelihood (ratio). For example, neural density estimation with autoregressive models and normalizing flows~\cite{larochelle2011neural,2015arXiv150505770J,papamakarios2017masked} have been used for this purpose and scale to higher dimensional data~\cite{Cranmer:2016lzt, 2018arXiv180507226P}. Alternatively, training a classifier to discriminate between $x \sim p(x|y)$ and $x \sim p(x|y')$ can be used to estimate the likelihood ratio $\hat{r}(x | y, y') \approx {p}(x|y)/{p}(x|y')$, which can be used for inference in either the frequentist or Bayesian paradigm~\cite{Cranmer:2015bka,Brehmer:2018hga, 2019arXiv190304057H}.

\begin{figure*}[htb]
    \centering
\includegraphics[width=\textwidth]{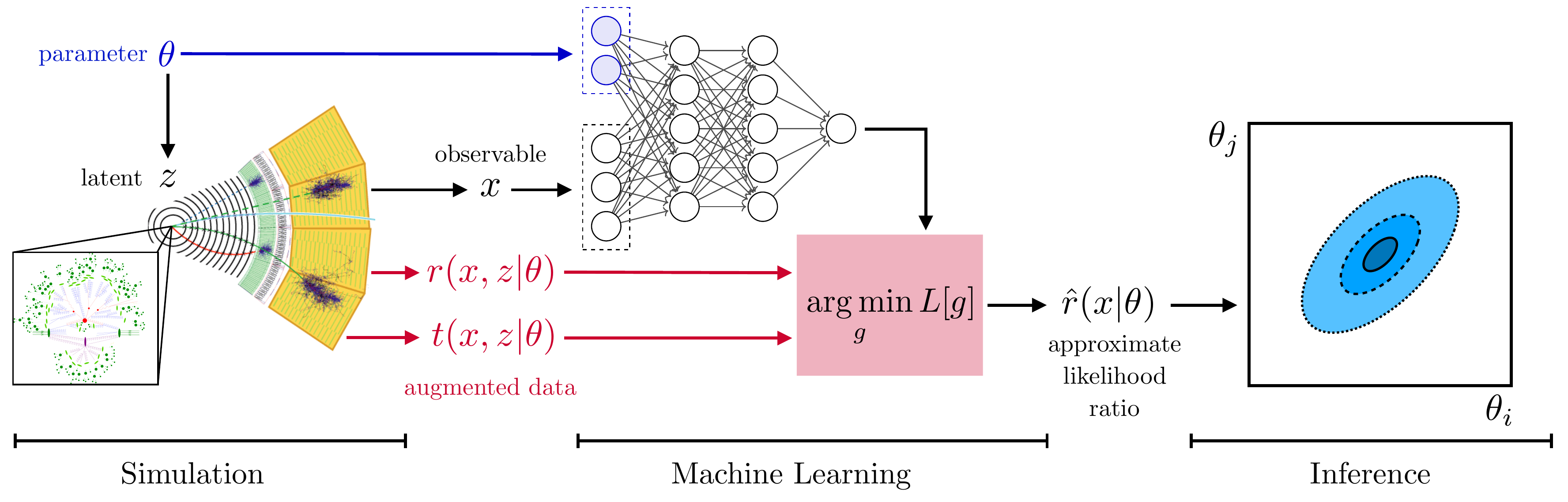}
    \caption{A schematic of machine learning based approaches to likelihood-free inference in which the simulation provides training data for a neural network that is subsequently used as a surrogate for the intractable likelihood during inference. Reproduced from ~\cite{Brehmer:2018kdj}.}
    \label{fig:likelihood-free}
\end{figure*}

\subsubsection{Examples in particle physics}

Thousands of published results within particle physics, including the discovery of the Higgs boson, involve statistical inference based on a surrogate likelihood $\hat{p}(x|y)$ constructed with density estimation techniques applied to synthetic datasets generated from the simulation.  These typically are restricted to one- or two-dimensional summary statistics or no features at all other than the number of events observed. While the term likelihood-free inference is relatively new, it is core to the methodology of experimental particle physics.

More recently, a suite of likelihood-free inference techniques based on neural networks have been developed and applied to models for physics beyond the standard model expressed in terms of effective field theory (EFT)~\cite{Brehmer:2018eca,Brehmer:2018kdj}. EFTs provide a systematic expansion of the theory around the standard model that is parametrized by coefficients for quantum mechanical operators, which play the role of $y$ in this setting. One interesting observation in this work is that even though the likelihood and likelihood ratio are intractable, the joint likelihood ratio $r(x,z|y, y')$ and the joint score $t(x,z|y) = \nabla_y \log p(x,z|y)$ are tractable and can be used to augment the training data (see Fig.~\ref{fig:likelihood-free}) and dramatically improve the sample efficiency of these techniques~\cite{Brehmer:2018hga}.

In addition, an \textit{inference compilation} technique has been applied to inference of a tau-lepton decay. This proof-of-concept effort required developing probabilistic programming protocol that can be integrated into existing domain-specific simulation codes such as \texttt{SHERPA} and \texttt{GEANT4}~\cite{Casado:2017cif,Baydin:2018npr}. This approach provides Bayesian inference on the latent variables $p(Z|X=x)$ and deep interpretability as the posterior corresponds to a distribution over complete stack-traces of the simulation, allowing any aspect of the simulation to be inspected probabilistically.



Another technique for likelihood-free inference that was motivated by the challenges of particle physics is known as adversarial variational optimization (AVO)~\cite{Louppe:2017pay}. AVO  parallels generative adversarial networks, where the generative model is no longer a neural network, but instead the domain-specific simulation. Instead of optimizing the parameters of the network, the goal is to optimize the parameters of the simulation so that the generated data matches the target data distribution. The main challenge is that, unlike neural networks, most scientific simulators are not differentiable. To get around this problem, a variational optimization technique is used, which provides a differentiable surrogate loss function. This technique is being investigated for tuning the parameters of the simulation, which is a computationally intensive task in which Bayesian optimization has also recently been used~\cite{Ilten:2016csi}.

\subsubsection{Examples in Cosmology}

Within Cosmology, early uses of ABC include constraining thick disk formation scenario of the Milky Way~\cite{2014A&A...569A..13R} and  inferences on rate of morphological transformation of galaxies at high redshift~\cite{2012MNRAS.425...44C}, which aimed to track the Hubble parameter evolution from type Ia supernova measurements. These experiences motivated the development of tools such as \texttt{CosmoABC} to streamline the application of the methodology in cosmological applications~\cite{2015AC....13....1I}.

More recently, likelihood-free inference methods based on machine learning have also been developed motivated by the experiences in cosmology. To confront the challenges of ABC for high-dimensional observations $X$, a data compression strategy was developed that learns summary statistics, that maximize the Fisher information on the parameters~\cite{Alsing:2018eau, Charnock:2018ogm}. The learned summary statistics approximate the sufficient statistics for the implicit likelihood in a small neighborhood of some nominal or fiducial parameter value. This approach is closely connected to that of \cite{Brehmer:2018hga}. Recently, these approaches have been extended to learn summary statistics that are robust to systematic uncertainties~\cite{Alsing:2019dvb}.


\subsection{Generative Models}
\begin{figure*}
    \centering
\includegraphics[width=\textwidth]{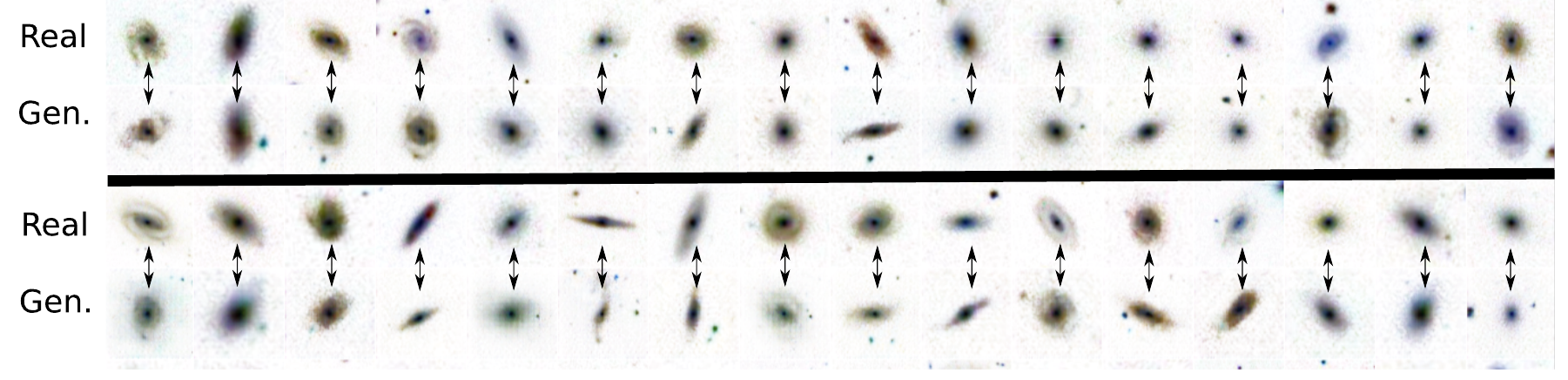}
    \caption{Samples from the GALAXY-ZOO dataset versus generated samples using conditional generative adversarial network. Each synthetic image is a 128$\times$128 colored image (here inverted) produced by conditioning on a set of features $y \in [0,1]^{37}$ . The pair of observed and generated images in each column correspond to the same $y$ value. Reproduced from ~\cite{2016arXiv160905796R}. }    \label{fig:galaxy-zoo}
\end{figure*}

An active area in machine learning research involves using unsupervised learning to train a generative model to produce a distribution that matches some empirical distribution. This includes generative adversarial networks (GANs)~\cite{goodfellow2014generative}, variational autoencoders (VAEs)~\cite{kingma2013auto, 2014arXiv1401.4082J}, autoregressive models, and models based on normalizing flows~\cite{larochelle2011neural,2015arXiv150505770J,papamakarios2017masked}.

Interestingly, the same issue that motivates likelihood-free inference, the intractability of the density implicitly defined by the simulator also appears in generative adversarial networks (GANs). If the density of a GAN were tractable, GANs would be trained via standard maximum likelihood, but because their density is intractable a trick was needed. The trick is to introduce an adversary -- i.e. the discriminator network used to classify the samples from the generative model and samples taken from the target distribution. The discriminator is effectively estimating the likelihood ratio between the two distributions, which provides a direct connections to the approaches to likelihood-free inference based on classifiers~\cite{Cranmer:2016lzt}.

Operationally, these models play a similar role as traditional scientific simulators, though traditional simulation codes also provide a causal model for the underlying data generation process grounded in physical principles. However, traditional scientific simulators are often very slow as the distributions of interest emerge from a low-level microphysical description. For example, simulating collisions at the LHC involves atomic-level physics of ionization and scintillation. Similarly,  simulations in cosmology involve gravitational interactions among enormous numbers of massive objects and may also include complex feedback processes that involve radiation, star formation, etc. Therefore, learning a fast approximation to these simulations is of great value.

Within particle physics early work in this direction included GANs for energy deposits from particles in calorimeters~\cite{Paganini:2017hrr, Paganini:2017dwg}, which is being studied by the ATLAS collaboration~\cite{ATL-SOFT-PUB-2018-001}.  In Cosmology, generative models have been used to learn the simulation for cosmological structure formation~\cite{2018ComAC...5....4R}. In an interesting hybrid approach, a deep neural network was used  to predict the non-linear structure formation of the universe from as a residual from a fast physical simulation based on linear perturbation theory~\cite{2018arXiv181106533H}.



In other cases, well-motivated simulations do not always exist or are impractical. Nevertheless, having a generative model for such data can be valuable for the purpose of calibration. An illustrative example in this direction comes from \cite{2016arXiv160905796R}, see Fig.~\ref{fig:galaxy-zoo}. The authors point out that the next generation of cosmological surveys for weak gravitational lensing rely on accurate measurements of the apparent shapes of distant galaxies. However, shape measurement methods require a precise calibration to meet the accuracy requirements of the science analysis. This calibration process is challenging as it requires large sets of high quality galaxy images, which are expensive to collect.  Therefore, the GAN enables an implicit generalization of the parametric bootstrap.

\subsection{Outlook and Challenges}\label{S:hep-outlook}

While particle physics and cosmology have a long history in utilizing machine learning methods, the scope of topics that machine learning is being applied to has grown significantly. Machine learning is now seen as a key strategy to confronting the challenges of the upgraded High-Luminosity LHC~\cite{Apollinari:2017cqg,Albertsson:2018maf} and is influencing the strategies for future experiments in both cosmology and particle physics~\cite{Ntampaka:2019udw}. One area in particular that has gathered a great deal of attention at the LHC is the challenge of identifying the tracks left by charged particles in high-luminostiy environments~\cite{Farrell:2018cjr}, which has been the focus of a recent kaggle challenge.

In almost all areas where machine learning is being applied to physics problems, there is a desire to incorporate domain knowledge in the form of hierarchical structure, compositional structure, geometrical structure, or symmetries that are known to exist in the data or the data-generation process. Recently, there has been a spate of work from the machine learning community in this direction~\cite{cohen2016group,
cohen2018spherical,2019arXiv190204615C,2018arXiv180609231K,2018arXiv180203690K,DBLP:journals/corr/abs-1803-01588,Bronstein:2016thv}. These developments are being followed closely by physicists, and already being incorporated into contemporary research in this area.

\section{Many-Body Quantum Matter}
\label{sec:many_body}

The intrinsic probabilistic nature of quantum mechanics makes physical
systems in this realm an effectively infinite source of \emph{big
data}, and a very appealing playground for ML applications. Paradigmatic
example of this probabilistic nature is the measurement process in
quantum physics. Measuring the \emph{position} $\mathbf{r}$ of an
electron orbiting around the nucleus can only be \emph{approximately
inferred} from measurements. An infinitely precise classical measurement
device can only be used to record the outcome of a specific observation
of the electron position. Ultimately, a complete characterization
of the measurement process is given by the wave function $\Psi(\mathbf{r})$,
whose square modulus ultimately defines the probability $P(\mathbf{r})=|\Psi(\mathbf{r})|^{2}$
of observing the electron at a given position in space. While in the
case of a single electron both theoretical predictions and experimental
inference for $P(\mathbf{r})$ are efficiently performed, the situation
becomes dramatically more complex in the case of many quantum particles.
For example, the probability of observing the positions of $N$ electrons
$P(\mathbf{r}_{1},\dots\mathbf{r}_{N})$ is an intrinsically high-dimensional
function, that can seldom be exactly determined for $N$ much larger
than a few tens. The exponential hardness in estimating $P(\mathbf{r}_{1},\dots\mathbf{r}_{N})$
is itself a direct consequence of estimating the complex-valued many-body
amplitudes $\Psi(\mathbf{r}_{1}\dots\mathbf{r}_{N})$ and is commonly
referred to as the quantum many-body problem. The quantum many-body
problem manifests itself in a variety of cases. These most chiefly include the theoretical
modeling and simulation of complex quantum systems -- most materials
and molecules -- for which only approximate solutions are often available.
Other very important manifestations of the quantum many-body problem include the understanding and analysis of experimental outcomes,
especially in relation with complex phases of matter.
In the following, we discuss some of the ML applications focused on
alleviating some of the challenging theoretical and experimental problems
posed by the quantum many-body problem.

\subsection{Neural-Network quantum states }
\label{sec:NNQS}

Neural-network quantum states (NQS) are a representation of the many-body
wave-function in terms of artificial neural networks (ANNs) \citep{carleo_solving_2017}.
A commonly adopted choice is to parameterize
wave-function amplitudes as a feed-forward neural network:
\begin{equation}
\Psi(\mathbf{r})=g^{(L)}(W^{(L)}  \dots g^{(2)} (W^{(2)} g^{(1)}(W^{(1)} \mathbf{r} ))), \label{eq:nqs}
\end{equation}
with similar notation to what introduced in Eq.~(\ref{eq:ffnn}).

\begin{figure}
    \centering
    \includegraphics[width=\columnwidth]{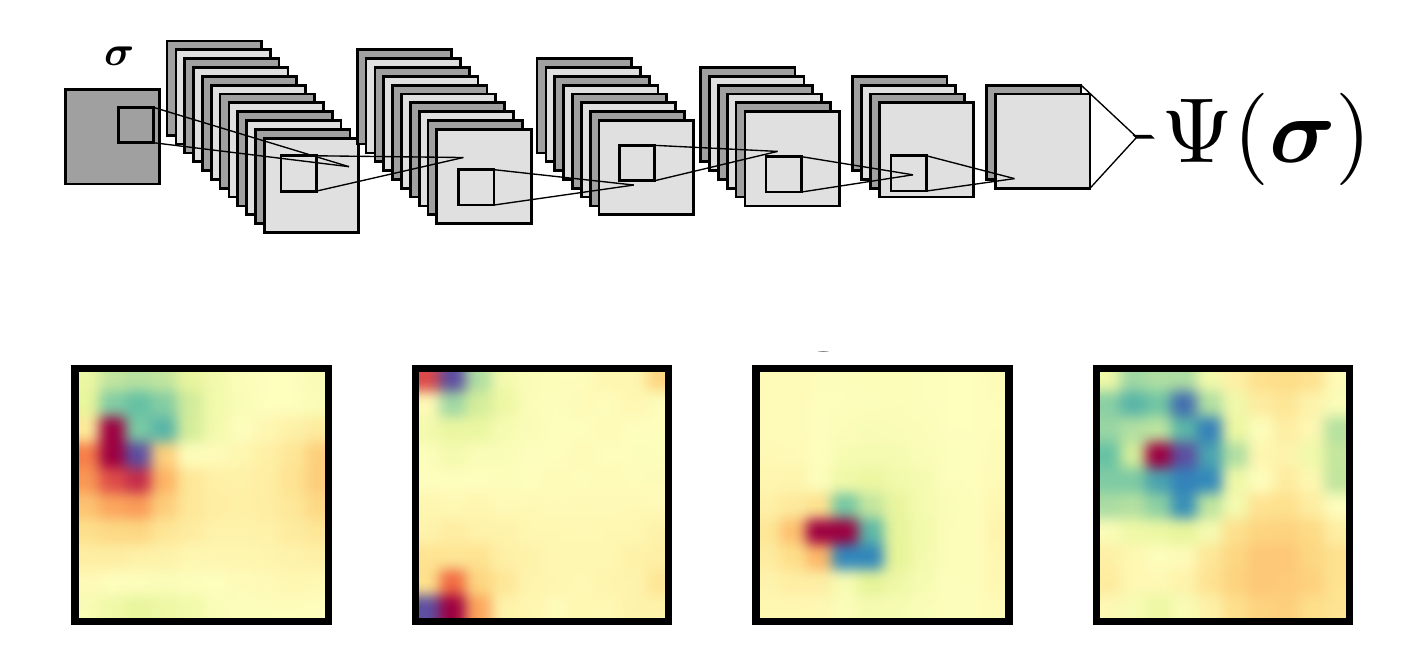}
    \caption{(Top) Example of a shallow convolutional neural network used to represent the many-body wave-function of a system of spin $1/2$ particles on a square lattice. (Bottom) Filters of a fully-connected convolutional RBM found in the variational learning of the ground-state of the two-dimensional Heisenberg model, adapted from \cite{carleo_solving_2017}. }
    \label{fig:nqs}
\end{figure}

Early works have mostly concentrated on shallow networks,
and most notably Restricted Boltzmann Machines (RBM)
\citep{Smolensky:1986:IPD:104279.104290}.
RBMs with hidden unit in $\{\pm1\}$ and without biases on the visible units formally correspond to FFNNs of depth $L=2$, and activations $g^{(1)}(x)=\log \cosh(x)$, $g^{(2)}(x)=\exp(x)$. An important difference with respect to RBM applications for
unsupervised learning of probability distributions, is that when used as NQS
RBM states are typically taken to have complex-valued weights \citep{carleo_solving_2017}.
Deeper architectures have been consistently studied and introduced
in more recent work, for example NQS based on fully-connected, and convolutional deep networks
\citep{saito_method_2018,choo_symmetries_2018,sharir_deep_2019}, see Fig.~\ref{fig:nqs} for a schematic example.
A motivation to use deep FFNN networks, apart from the practical success
of deep learning in industrial applications, also comes from more general theoretical arguments
in quantum physics. For example, it has been shown that deep NQS can sustain entanglement more efficiently
than RBM states \citep{levine_quantum_2019,liu_machine_2017}.
Other extensions of the NQS representation concern representation of
mixed states described by density matrices, rather than pure wave-functions.
In this context, it is possible to define positive-definite RBM parametrizations of the density matrix \citep{torlai_latent_2018}.

One of the specific challenges emerging in the quantum domain is imposing
physical symmetries in the NQS representations. In the case of a periodic
arrangement of matter, spatial symmetries can be imposed using convolutional
architectures similar to what is used in image classification tasks \citep{saito_method_2018,choo_symmetries_2018,sharir_deep_2019}.
Selecting high-energy states in different symmetry sectors has also
been demonstrated \citep{choo_symmetries_2018}. While spatial symmetries
have analogous counterparts in other ML applications, satisfying more
involved quantum symmetries often needs a deep rethinking of ANN architectures.
The most notable case in this sense is the \emph{exchange symmetry}.
For bosons, this amounts to imposing the wave-function to be permutationally
invariant with respect to exchange of particle indices. The Bose-Hubbard
model has been adopted as a benchmark for ANN bosonic architectures,
with state-of-the-art results having been obtained \citep{saito_machine_2017,saito_solving_2017,saito_method_2018,teng_machine-learning_2018}.
The most challenging symmetry is, however, certainly the fermionic one.
In this case, the NQS representation needs to encode the antisymmetry
of the wave-function (exchanging two particle positions, for example,
leads to a minus sign). In this case, different approaches have been
explored, mostly expanding on existing variational ansatz for fermions.
A symmetric RBM wave-function correcting an antisymmetric correlator
part has been used to study two-dimensional interacting lattice fermions \citep{nomura_restricted_2017}.
Other approaches have tackled the fermionic symmetry problem using a backflow
transformation of Slater determinants \citep{luo_backflow_2018},
or directly working in first quantization \citep{han_solving_2018}.
The situation for fermions is certainly the most challenging for ML
approaches at the moment, owing to the specific nature of the symmetry.
On the applications side, NQS representations have been used so-far
along three main different research lines.

\subsubsection{Representation theory}

An active area of research concerns the general expressive power
of NQS, as also compared to other families of variational states.
Theoretical activity on the representation properties
of NQS seeks to understand how large, and how deep
should be neural networks describing interesting interacting quantum systems.
In connection with the first numerical results obtained with RBM states,
the entanglement has been soon identified as
a possible candidate for the expressive power of NQS. RBM states for example
can efficiently support volume-law scaling \citep{deng_quantum_2017}, with a number of
variational parameters scaling only polynomially with system size.
In this direction, the language of tensor networks has been particularly
helpful in clarifying some of the properties of NQS \citep{chen_equivalence_2018,pastori_generalized_2018}.
A family of NQS based on RBM states has been shown to be equivalent
to a certain family of variational states known as correlator-product-states
\citep{glasser_neural-network_2018,clark_unifying_2018}. The question
of determining how large are the respective classes of quantum states
belonging to the NQS form, Eq. (\ref{eq:nqs}) and to computationally efficient tensor network
is, however, still open. Exact representations of several
intriguing phases of matter, including topological states and stabilizer codes \citep{deng_machine_2017,glasser_neural-network_2018,kaubruegger_chiral_2018,lu_efficient_2018,huang_neural_2017,zheng_restricted_2018},
have also been obtained in closed RBM form. Not surprisingly, given its shallow depth,
RBM architectures are also expected to have limitations, on general grounds.
Specifically, it is not in general
possible to write \emph{all} possible physical states in terms of
compact RBM states \citep{gao_efficient_2017}. In order to lift the
intrinsic limitations of RBMs, and efficiently describe
a very large family of physical states, it is necessary to introduce
deep Boltzmann Machines (DBM) with two hidden layers \citep{gao_efficient_2017}.
Similar network constructions
have been introduced also as a possible theoretical framework, alternative
to the standard path-integral representation of quantum mechanics
\citep{carleo2018constructing}.

\subsubsection{Learning from data}

Parallel to the activity on understanding the theoretical properties of NQS,
a family of studies in this field is concerned with the
problem of understanding how hard it is, in practice, to learn
a quantum state from numerical data.
This can be realized using either synthetic data (for example coming from numerical simulations)
or directly from experiments.

This line of research has been explored in the supervised learning setting, to
understand how well NQS can represent states that are not easily expressed (in closed analytic form) as ANN.
The goal is then to train a NQS network $|\Psi\rangle$ to represent, as close as possible, a certain target state $|\Phi\rangle$
whose amplitudes can be efficiently computed.
This approach has been successfully used to learn ground-states of fermionic, frustrated, and bosonic Hamiltonians \citep{cai_approximating_2018}.
Those represent interesting study cases, since the sign/phase structure of the target wave-functions can pose a challenge to
standard activation functions used in FFNN.
Along the same lines, supervised approaches have been proposed to learn random matrix product states wave-functions  both with shallow NQS \cite{borin_approximating_2019}, and with generalized NQS including
a computationally treatable DBM form \cite{pastori_generalized_2018}.
While in the latter case these studies have revealed efficient strategies to
perform the learning, in the former case hardness in learning some random MPS has been showed.
At present, it is speculated that this hardness
originates from the entanglement structure of the random MPS, however it is unclear if this is related to the hardness of the NQS optimization landscape or to an intrinsic limitation of shallow NQS.

Besides supervised learning of given quantum states, data-driven approaches with NQS have largely concentrated on unsupervised approaches.
In this framework, only measurements from some target state $|\Phi\rangle$ or density matrix are available, and the goal is to reconstruct the full state, in NQS form, using such measurements.
In the simplest setting, one is given a data set of $M$ measurements $\mathbf{r}^{(1)}\dots \mathbf{r}^{(M)}$ distributed according to Born's rule prescription $P(\mathbf{r})=|\Phi(\mathbf{r})|^2$, where $P(\mathbf{r})$ is to be reconstructed.
In cases when the wave-function is positive definite, or when only measurements in a certain basis are provided, reconstructing $P(\mathbf{r})$ with standard unsupervised learning approaches is enough to reconstruct all the available information on the underlying quantum state $\Phi$.
This approach for example has been demonstrated for ground-states of stoquastic Hamiltonians \cite{torlai_neural-network_2018} using RBM-based generative models.
An approach based on deep VAE generative models has also been demonstrated in the case of a family of classically-hard to sample from quantum states \cite{rocchetto_learning_2018}, for which the effect of network depth has been shown to be beneficial for compression.

In the more general setting, the problem is to reconstruct a general quantum state, either pure or mixed, using measurements from more
than a single basis of quantum numbers. Those are especially necessary to reconstruct also the complex phases of the quantum state.
This problem corresponds to a well-known problem in quantum information, known as quantum state tomography, for which specific NQS approaches have been introduced \cite{torlai_neural-network_2018,torlai_latent_2018,carrasquilla_reconstructing_2018}. Those are discussed more in detail, in the dedicated section \ref{subsec:tomography}, also in connection with other ML techniques used for this task.

\subsubsection{Variational Learning}

Finally, one of the main applications for the NQS representations is in the context
of variational approximations for many-body quantum problems.
The goal of these approaches is, for example, to approximately solve the Schr\"{o}dinger equation using a NQS representation for the wave-function.
In this case, the problem of finding the ground state of a given quantum Hamiltonian $H$ is formulated in variational terms as the problem of learning NQS weights $W$
minimizing $E(W)=\langle\Psi(W)|H|\Psi(W)\rangle/\langle\Psi(W)|\Psi(W)\rangle$. This is achieved using a learning scheme based on variational
Monte Carlo optimization \cite{carleo_solving_2017}.
Within this family of applications,
no external data representative of the quantum state is given, thus they typically demand a larger computational burden than supervised and unsupervised learning schemes for NQS.

Experiments on a variety of spin \citep{glasser_neural-network_2018,choo_symmetries_2018,liang_solving_2018,deng_machine_2017}, bosonic \citep{saito_machine_2017,saito_solving_2017,saito_method_2018,choo_symmetries_2018},
and fermionic \citep{nomura_restricted_2017,luo_backflow_2018,han_solving_2018}
models have shown that results competitive with existing state-of-the-art approaches can be obtained.
In some cases, improvement over
existing variational results have been demonstrated, most notably
for two-dimensional lattice models \citep{carleo_solving_2017,nomura_restricted_2017,luo_backflow_2018}
and for topological phases of matter \cite{glasser_neural-network_2018,kaubruegger_chiral_2018}.

Other NQS applications concern the solution of the time-dependent
Schr\"{o}dinger equation \citep{carleo_solving_2017,czischek_quenches_2018,schmitt_quantum_2018,fabiani_investigating_2019}.
In these applications, one uses the time-dependent
variational principle of Dirac and Frenkel \citep{dirac_note_1930,frenkel_wave_1934}
to learn the optimal time evolution of network weights.
This can be suitably generalized also to open dissipative quantum systems,
for which a variational solution of the Lindblad equation can be realized  \citep{hartmann_neural-network_2019,yoshioka_constructing_2019,nagy_variational_2019,vicentini_variational_2019}.

In the great majority of the variational applications discussed here, the learning schemes used are typically higher-order techniques than
standard SGD approaches. The stochastic reconfiguration
(SR) approach \citep{sorella_green_1998,becca_quantum_2017} and its generalization to the time-dependent case \citep{carleo_localization_2012},
have proven particularly
suitable to variational learning of NQS. The SR scheme can be seen
as a quantum analogous of the natural-gradient method for learning
probability distributions \citep{amari_natural_1998}, and builds
on the intrinsic geometry associated with the neural-network parameters.
More recently, in an effort to use deeper and more expressive networks than those initially adopted,
learning schemes building on first-order techniques have been more consistently used \cite{kochkov_variational_2018,sharir_deep_2019}.
These constitute two different philosophy of approaching the same problem.
On one hand, early applications focused on small networks learned with very accurate but expensive training techniques.
On the other hand, later approaches have focused on deeper networks and cheaper --but also less accurate-- learning techniques.
Combining the two philosophy in a computationally efficient way is one of the open challenges in the field.

\subsection{Speed up many-body simulations }

The use of ML methods in the realm of the quantum many-body problems
extends well beyond neural-network representation of quantum states.
A powerful technique to study interacting models are Quantum Monte
Carlo (QMC) approaches. These methods stochastically compute properties
of quantum systems through mapping to an effective classical model,
for example by means of the path-integral representation. A practical
issue often resulting from these mappings is that providing efficient
sampling schemes of high-dimensional spaces (path integrals, perturbation
series, etc..) requires a careful tuning, often problem-dependent.
Devising general-purpose samplers for these representations is therefore
a particularly challenging problem. Unsupervised ML methods can, however,
be adopted as a tool to speed-up Monte Carlo sampling for both classical and
quantum applications.
Several approaches in this direction have been proposed, and leverage the ability of
unsupervised learning to well approximate the target
distribution being sampled from in the underlying Monte Carlo scheme.
Relatively simple energy-based generative models
have been used in early applications for classical systems
\citep{huang_accelerated_2017,liu_self-learning_2017}. "Self-learning" Monte Carlo
techniques have then been generalized also to fermionic systems
\citep{liu_self-learning_fermions_2017,nagai_self-learning_2017,chen_symmetry-enforced_2018}.
Overall, it has been found that such approaches are effective at
reducing the autocorrelation times, especially when compared to families of less
effective Markov Chain Monte Carlo with local updates.
More recently, state-of-the-art generative ML models have been adopted to speed-up sampling
in specific tasks. Notably, \citep{wu2018solving} have used deep autoregressive models
that may enable a more efficient sampling from hard classical problems, such as
spin glasses.
The problem of finding efficient sampling schemes for the underlying
classical models is then transformed into the problem of finding an efficient corresponding
autoregressive deep network representation.
This approach has also been generalized to the quantum cases in \cite{sharir_deep_2019},
where an autoregressive representation of the wave-function is introduced.
This representation is automatically normalized and allows to bypass Markov Chain Monte Carlo in the variational learning discussed above.

While exact for a large family of bosonic and spin systems, QMC techniques
typically incur in a severe sign problem when dealing with several
interesting fermionic models, as well as frustrated spin Hamiltonians.
In this case, it is tempting to use ML approaches to attempt a direct
or indirect reduction of the sign problem. While only in its first
stages, this family of applications has been used to infer information
about fermionic phases through hidden information in the Green's function
\citep{broecker_machine_2017}.

Similarly, ML techniques can help reduce the burden of more subtle
manifestations of the sign problem in dynamical properties of quantum
models. In particular, the problem of reconstructing spectral functions
from imaginary-time correlations in imaginary time is also a field
in which ML can be used as an alternative to traditional maximum-entropy
techniques to perform analytical continuations of QMC data \citep{arsenault_projected_2017,yoon_analytic_2018,fournier_artificial_2018}.

\subsection{Classifying many-body quantum phases}
\label{subsec-class_manybody}

The challenge posed by the complexity of many-body quantum states
manifests itself in many other forms. Specifically, several elusive phases of quantum matter are often hard to characterize and pinpoint both in numerical simulations and in experiments. For this reason, ML schemes to identify phases of matter have become particularly popular in the context of quantum phases. In the following we review some of the specific applications to the quantum domain, while a more general discussion on identifying phases and phase transitions is to be found in \ref{subsec:ml_in_stat_phys}.

\subsubsection{Synthetic data}
Following the early developments in phase classifications with supervised approaches \cite{carrasquilla2017machine,wang_discovering_2016,van2017learning}, many studies have since then focused on analyzing phases of matter in synthetic data, mostly from simulations of quantum systems. While we do not attempt here to provide an exhaustive review of the many studies appeared in this direction, we highlight two large families of problems that have so-far largely served as benchmarks for new ML tools in the field.

A first challenging test bench for phase classification schemes is the case of quantum many-body localization. This is an elusive phase of matter showing characteristic fingerprints in the many-body wave-function itself, but not necessarily emerging from more traditional order parameters [see for example \cite{alet_many-body_2018} for a recent review on the topic]. First studies in this direction have focused on training strategies aiming at the Hamiltonian or entanglement spectra \cite{schindler2017probing,venderley_machine_2018,hsu_machine_2018,huembeli_automated_2018,zhang_interpretable_2019}. These works have demonstrated the ability to very effectively learn the MBL phase transition in relatively small systems accessible with exact diagonalization techniques. Other studies have instead focused on identifying signatures directly in experimentally relevant quantities, most notably from the many-body dynamics of local quantities \cite{van_nieuwenburg_learning_2018,doggen_many-body_2018}. The latter schemes appear to be at present the most promising for applications to experiments, while the former have been used as a tool to identify the existence of an unexpected phase in the presence of correlated disorder \cite{hsu_machine_2018}.

Another very challenging class of problems is found when analyzing topological phases of matter. These are largely considered a non-trivial test for ML schemes, because these phases are typically characterized by non-local order parameters. In turn, these non-local order parameters are hard to learn for popular classification schemes used for images.
This specific issue is already present when analyzing classical models featuring topological phase transitions. For example, in the presence of a BKT-type transition, learning schemes trained on raw Monte Carlo configurations are not effective \cite{beach_machine_2018,hu_discovering_2017}. These problems can be circumvented devising training strategies using pre-engineered features \cite{wetzel2017unsupervised,wang2017machine,cristoforetti_towards_2017,broecker_quantum_2017} instead of raw Monte Carlo samples. These features typically rely on some important a-priori assumptions on the nature of the phase transition to be looked for, thus diminishing their effectiveness when looking for new phases of matter.
Deeper in the quantum world, there has been research activity along the direction of learning, in a supervised fashion, topological invariants.  Neural networks can be used for example to classify families of non-interacting topological Hamiltonians, using as an input their discretized coefficients, either in real  \cite{ohtsuki_deep_2016,ohtsuki_deep_2017} or momentum space \cite{sun_deep_2018,zhang_machine_2018}. In these  cases, it is found that neural networks are able to reproduce the (already known beforehand) topological invariants, such as winding numbers, Berry curvatures and more.
The context of strongly-correlated topological matter is, to a large extent, more challenging than the case of non-interacting band models. In this case, a common approach is to define a set of carefully pre-engineered features to be used on top of the raw data. One well known example is the case of of the so-called \emph{quantum loop topography} \cite{zhang_quantum_2017}, trained on local operators computed on single shots of sampled wave-function walkers, as for example done in variational Monte Carlo. It has been shown that this very specific choice of local features is able to distinguish strongly interacting fraction Chern insulators, and also $\mathbb{Z}_{2}$ quantum spin liquids \citep{zhang_machine_2017}. Similar efforts have been realized to classify more exotic phases of matter, including magnetic skyrmion phases \cite{iakovlev2018supervised}, and dynamical states in antiskyrmion dynamics \cite{ritzmann_trochoidal_2018}.

Despite the progress seen so far along the many direction described here, it is fair to say that topological phases of matter, especially for interacting systems, constitute one of the main challenges for phase classification. While some good progress has already been made \cite{rodriguez-nieva_identifying_2018,huembeli_identifying_2018}, future research will need to address the issue of finding training schemes not relying on pre-selection of data features.

\subsubsection{Experimental data}
Beyond extensive studies on data from numerical simulations, supervised schemes have found their way also as a tool to analyze experimental data from quantum systems. In ultra-cold atoms experiments, supervised learning tools have been used to map out both the topological phases of non-interacting particles, as well the onset of Mott insulating phases in finite optical traps \cite{rem_identifying_2018}.
In this specific case, the phases where already known and identifiable with other approaches. However, ML-based techniques combining a-priori theoretical knowledge with experimental data hold the potential for genuine scientific discovery.

\begin{figure}
    \centering
   \includegraphics[width=\columnwidth]{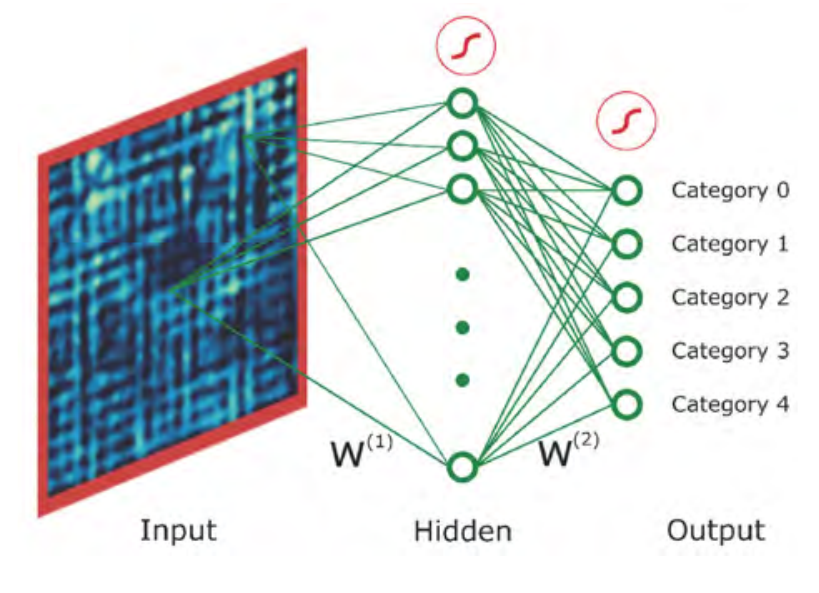}
    \caption{Example of machine learning approach to the classification of experimental images from scanning tunneling microscopy of high-temperature superconductors. Images are classified according to the predictions of distinct types of periodic spatial modulations. Reproduced from \cite{zhang_using_2018} }
    \label{fig:quantum_classification}
\end{figure}

For example, ML can enable scientific discovery in the interesting cases when experimental data has to be attributed to one of many available and equally likely a-priory theoretical models, but the experimental information at hand is not easily interpreted. Typically interesting cases emerge for example when the order parameter is a complex, and only implicitly known, non-linear function of the experimental outcomes. In this situation, ML approaches can be used as a powerful tool to
effectively learn the underlying traits of a given theory, and provide a possibly unbiased classification of experimental data.
This is the case for incommensurate phases in high-temperature  superconductors, for which scanning tunneling microscopy images
reveal complex patters that are hard to decipher using conventional analysis tools. Using supervised approaches in this context, recent work \cite{zhang_using_2018} has shown that is possible to infer the nature of spatial ordering in these systems, also see Fig.~\ref{fig:quantum_classification}.

A similar idea has been also used for another prototypical interacting quantum systems of fermions, the Hubbard model, as implemented in ultra-cold atoms experiments in optical lattices. In this case the reference models provide snapshots of the thermal density matrix that can be pre-classified in a supervised learning fashion. The outcome of this study \cite{bohrdt_classifying_2018}, is that the experimental results are with good confidence compatible with one of the theories proposed, in this case a geometric string theory for charge carriers.

In the last two experimental applications described above, the outcome of the supervised approaches are to a large extent highly non-trivial, and hard to predict a priori on the basis of other information at hand. The inner bias induced by the choice of the theories to be classified is however one of the current limitations that these kind of approaches face.

\subsection{Tensor networks for machine learning}
The research topics reviewed so far are mainly concerned with the
use of ML ideas and tools to study problems in the realm of quantum many-body physics.
Complementary to this philosophy, an interesting research direction
in the field explores the inverse direction, investigating how ideas
from quantum many-body physics can inspire and devise new powerful ML tools.
Central to these developments are tensor-network representations of many-body quantum states.
These are very successful variational families of many-body wave functions, naturally
emerging from low-entanglement representations of quantum states \cite{verstraete_matrix_2008}.
Tensor networks can serve both as a practical and a conceptual tool for ML tasks,
both in the supervised and in the unsupervised setting.

These approaches build on the idea of providing physics-inspired learning schemes and
network structures alternative to the more conventionally adopted stochastic learning
schemes and FFNN networks. For example, matrix product state (MPS) representations,
a work-horse for the simulation of interacting one-dimensional quantum systems
\cite{white_density_1992}, have been re-purposed to perform classification tasks,
\cite{novikov_exponential_2016,stoudenmire_supervised_2016,liu_entanglement-guided_2018}, and also recently adopted as
explicit generative models for unsupervised learning \citep{han_unsupervised_2018,stokes2019probabilistic}.
It is worth mentioning that other related high-order tensor decompositions, developed in the context of applied mathematics have been used for ML purposes
\citep{acar_unsupervised_2009,anandkumar_tensor_2014}.
Tensor-train decompositions \cite{oseledets_tensor-train_2011}, formally equivalent to MPS
representations, have been introduced in parallel as a tool to perform
various machine learning tasks \cite{novikov_exponential_2016,izmailov_2017,gorodetsky_2018}.
Networks closely related to MPS have also been explored for time-series modeling
\cite{guo_matrix_2018}.

In the effort of increasing the amount of entanglement encoded in these low-rank tensor
decompositions, recent works have concentrated on tensor-network representations alternative
to the MPS form. One notable example is the use of tree tensor networks with a hierarchical
structure \cite{shi_classical_2006,hackbusch_new_2009}, which have been applied to classification \cite{liu_machine_2017,stoudenmire_learning_2018}
and generative modeling \cite{cheng_tree_2019} tasks with good success.
Another example is the use of entangled plaquette
states \cite{gendiar_latent_2002,mezzacapo_ground-state_2009,changlani_approximating_2009}
and string bond states \cite{schuch_simulation_2008}, both showing sizable improvements
in classification tasks over MPS states \cite{glasser_supervised_2018}.

On the more theoretical side, the deep connection between tensor networks and complexity
measures of quantum many-body wave-functions, such as entanglement entropy, can be used to
understand, and possible inspire, successful network designs for ML purposes.
The tensor-network formalism has proven powerful in interpreting deep learning through the
lens of renormalization group concepts. Pioneering work in this direction has connected MERA
tensor network states \cite{vidal_entanglement_2007} to hierarchical Bayesian networks \cite{beny2013deep}.
In later analysis, convolutional
arithmetic circuits \cite{pmlr-v49-cohen16}, a family of convolutional networks with
product non-linearities, have been introduced as a convenient model to bridge tensor
decompositions with FFNN architectures.
Beside their conceptual relevance, these connections can help clarify the role of
\emph{inductive bias} in modern and commonly adopted neural networks \cite{levine_deep_2017}.

\subsection{Outlook and Challenges}

Applications of ML to quantum many-body problems have seen a fast-pace
progress in the past few years, touching a diverse selection of topics ranging from
numerical simulation to data analysis. The potential of ML techniques has
already surfaced in this context, already showing improved performance with respect to existing techniques
on selected problems. To a large extent, however, the real power of
ML techniques in this domain has been only partially demonstrated, and several open problems remain to be addressed.

In the context of variational studies with NQS, for example, the origin of the empirical success obtained so far with different kind of neural network quantum states is not equally well understood as for other families of variational states, like tensor networks. Key open challenges remain also with the representation and simulation of fermionic systems, for which efficient neural-network representation are still to be found.

Tensor-network representations for ML purposes, as well as complex-valued networks like those used for NQS, play an important role to bridge the field back to the arena of computer science. Challenges for the future of this research direction consist in effectively interfacing with
the computer-science community, while retaining the interests and the generality of the physics tools.

For what concerns ML approaches to experimental data, the field is largely still in its infancy, with only a few applications having been demonstrated so far. This is in stark contrast with other fields, such as High-Energy and Astrophysics, in which ML approaches have matured to a stage where they are often used as standard tools for data analysis. Moving towards achieving the same goal in the quantum domain demands closer collaborations between the theoretical and experimental efforts, as well as a deeper understanding of the specific problems where ML can make a substantial difference.

Overall, given the relatively short time span in which applications of ML approaches to many-body quantum matter have emerged, there are however good reasons to believe that these challenges will be energetically addressed--and some of them solved-- in the coming years.

\section{Quantum computing}
\label{sec:quantum}

Quantum computing uses quantum systems to process information. In the most popular framework of gate-based quantum computing \cite{nielsen2002quantum}, a quantum algorithm describes the evolution of an initial state  $|\psi_0 \rangle$ of a quantum system of $n$ two-level systems called \textit{qubits} to a final state $| \psi_f \rangle$ through discrete transformations or \textit{quantum gates}. The gates usually act only on a small number of qubits, and the sequence of gates defines the computation.

The intersection of machine learning and quantum computing has become an active research area in the last couple of years, and contains a variety of ways to merge the two disciplines (see also \cite{dunjko2018machine} for a review). \textit{Quantum machine learning} asks how quantum computers can enhance, speed up or innovate machine learning \citep{biamonte17, schuld18quantum,ciliberto2018quantum} (see also Sections \ref{sec:instruments} and \ref{sec:quantum}). \textit{Quantum learning theory} highlights theoretical aspects of learning under a quantum framework \citep{arunachalam2017guest}.

In this Section we are concerned with a third angle, namely how machine learning can help us to build and study quantum computers. This angle includes topics ranging from the use of intelligent data mining methods to find physical regimes in materials that can be used as qubits \citep{kalantre2019machine}, to the verification of quantum devices \citep{agresti2019pattern}, learning the design of quantum algorithms \citep{bang2014strategy, wecker2016training}, facilitating classical simulations of quantum circuits \citep{jonsson_neural-network_2018}, automated design on quantum experiments \cite{krenn2016automated,melnikov2018active}, and learning to extract relevant information from measurements \citep{seif_machine_2018}.

We focus on three general problems related to quantum computing which were targeted by a range of ML methods: the problem of reconstructing en benchmarking quantum states via measurements; the problem of preparing a quantum state via \textit{quantum control}; the problem of maintaining the information stored in the state through \textit{quantum error correction}. The first problem is known as quantum state tomography, and it is especially useful to understand and improve upon the limitations of current quantum hardware. Quantum control and quantum error corrections solve related problems, however usually the former refers to hardware-related solutions while the latter uses algorithmic solutions to the problem of executing a computational protocol with a quantum system.

Similar to the other disciplines in this review, machine learning has shown promising results in all these areas, and will in the longer run likely enter the toolbox of quantum computing to be used side-by-side with other well-established methods.

\subsection{Quantum state tomography}
\label{subsec:tomography}
The general goal of quantum state tomography (QST) is to reconstruct the density matrix of an unknown quantum state, through experimentally
available measurements. QST is a central tool in several fields of quantum information and quantum technologies in general, where it is often used
as a way to assess the quality and the limitations of the experimental platforms. The resources needed to perform full QST are however extremely
demanding, and the number of required measurements scales exponentially with the number of qubits/quantum degrees of freedom [see \cite{paris_quantum_2004} for a review on the topic, and \cite{haah2017sample, o2016efficient} for a discussion on the hardness of learning in state tomography].

ML tools have been identified already several years ago as a tool to improve upon the cost of full QST, exploiting some special structure in the density matrix. Compressed sensing \cite{gross_quantum_2010} is one prominent approach to the problem, allowing to reduce the number of required measurements from $d^2$ to $\mathcal{O}(rd \log(d)^2)$, for a density matrix of rank $r$ and dimension $d$. Successful experimental realization of this technique has been for example implemented for a six-photon state \cite{toth2010permutationally} or a seven-qubit system of trapped ions \cite{riofrio_experimental_2017}.
On the methodology side, full QST has more recently seen the development of deep learning approaches. For example, using a supervised approach based on neural networks having as an output the full density matrix, and as an input possible measurement outcomes \cite{xu_neural_2018}. The problem of choosing optimal measurement basis for QST has also been recently addressed using a neural-network based approach that optimizes the prior distribution on the target density matrix, using Bayes rule \cite{quek_adaptive_2018}.
In general, while ML approaches to full QST can serve as a viable tool to alleviate the measurement requirements, they cannot however provide an improvement over the intrinsic exponential scaling of QST.

The exponential barrier can be typically overcome only in situations when the quantum state is assumed to have some specific regularity properties.
Tomography based on tensor-network paremeterizations of the density matrix has been an important first step in this direction, allowing for tomography of large, low-entangled quantum systems \cite{lanyon_efficient_2017}.
ML approaches to parameterization-based QST have emerged in recent times as a viable alternative, especially for highly entangled states. Specifically, assuming a NQS form (see Eq. \ref{eq:nqs} in the case of pure states) QST can be
reformulated as an unsupervised ML learning task. A scheme to retrieve the phase of the wave-function, in the case of pure states,
has been demonstrated in \citep{torlai_neural-network_2018}.
In these applications, the complex phase of the many-body wave-function
is retrieved upon reconstruction of several probability densities
associated to the measurement process in different basis. Overall, this approach has allowed to demonstrate QST of highly entangled states up to about 100 qubits, unfeasible for full QST techniques. This tomography approach can
be suitably generalized to the case of mixed states introducing parameterizations
of the density matrix based either on purified NQS \citep{torlai_latent_2018} or on deep normalizing flows \cite{cranmer_inferring_2019}.
The former approach has been also demonstrated experimentally with Rydberg atoms \cite{torlai_integrating_2019}.
An interesting alternative to the NQS representation for tomographic purposes
has also been recently suggested \citep{carrasquilla_reconstructing_2018}.
This is based on parameterizing the density matrix
directly in terms of positive-operator valued measure (POVM) operators.
This approach therefore has the important advantage of directly learning the measurement process itself,
and has been demonstrated to scale well on rather large mixed states.
A possible inconvenient of this approach is that the density matrix is only implicitly defined in terms of generative models, as opposed to explicit parameterizations found in NQS-based approaches.

Other approaches to QST have explored the use of quantum states parameterized as ground-states of local Hamiltonians \cite{xin_local-measurement-based_2018}, or the intriguing possibility of bypassing QST to directly measure quantum entanglement \cite{gray_machine-learning-assisted_2018}. Extensions to the more complex problem of quantum process tomography are also promising \cite{banchi_modelling_2018}, while the scalability of ML-based approaches to larger systems still presents challenges.

Finally, the problem of learning quantum states from experimental measurements has also profound implications on the understanding of the complexity of quantum systems.
In this framework, the PAC learnability of quantum states \cite{aaronson2007learnability}, experimentally demonstrated in \cite{rocchetto_experimental_2017}, and the ‘’shadow tomography” approach \cite{aaronson_shadow_2017}, showed that even linearly sized training sets can provide sufficient information to succeed in certain quantum learning tasks. These information-theoretic guarantees come with computational restrictions and learning is efficient only for special classes of states \cite{rocchetto2018stabiliser}

\subsection{Controlling and preparing qubits}

A central task of quantum control is the following: Given an evolution $U(\theta)$ that depends on parameters $\theta$ and maps an initial quantum state $| \psi_0 \rangle$ to $| \psi(\theta)\rangle = U(\theta) | \psi_0 \rangle$, which parameters $\theta^*$ minimise the overlap or distance between the prepared state and the target state, $|\langle \psi(\theta)| \psi_{\text{target}}\rangle|^2$? To facilitate analytic studies, the space of possible control interventions is often discretized, so that $U(\theta) = U(s_1,\dots,s_T)$ becomes a sequence of steps $s_1,\dots,s_T$. For example, a control field could be applied at only two different strengths $h_1$ and $h_2$, and the goal is to find an optimal strategy $s_t \in \{h_1, h_2\}$, $t=1,\dots,T$ to bring the initial state as close as possible to the target state using only these discrete actions.

This setup directly generalizes to a reinforcement learning framework \cite{sutton2018reinforcement}, where an agent picks ``moves'' from the list of allowed control interventions, such as the two field strengths applied to the quantum state of a qubit. This framework has proven to be competitive to state-of-the-art methods in various settings, such as state preparation in non-integrable  many-body  quantum  systems  of  interacting  qubits \citep{bukov_reinforcement_2018}, or the use of strong periodic oscillations to prepare so-called ``Floquet-engineered'' states \citep{bukov_floquet_2018}. A recent study comparing (deep) reinforcement learning with traditional optimization methods such as Stochastic Gradient Descent for the preparation of a single qubit state shows that learning is of advantage if the ``action space'' is naturally discretized and sufficiently small \citep{zhang2019reinforcement}.

The picture becomes increasingly complex in slightly more realistic settings, for example when the control is noisy \citep{niu2018universal}. In an interesting twist, the control problem has also been tackled by predicting future noise using a recurrent neural network that analyses the time series of past noise. Using the prediction, the anticipated future noise can be corrected \citep{mavadia_prediction_2017}.

An altogether different approach to state preparation with machine learning tries to find optimal strategies for evaporative cooling to create Bose-Einstein condensates \citep{wigley_fast_2016}. In this online optimization strategy based on Bayesian optimization~\cite{Jones1998,frazier2018tutorial}, a Gaussian process is used as a statistical model that captures the relationship between the control parameters and the quality of the condensate. The strategy discovered by the machine learning model allows for a cooling protocol that uses $10$ times fewer iterations than pure optimization techniques. An interesting feature is that - contrary to the common reputation of machine learning - the Gaussian process allows to determine which control parameters are more important than others.

Another angle is captured by approaches that `learn' the sequence of optical instruments in order to prepare highly entangled photonic quantum states \cite{melnikov2018active}.

\subsection{Error correction}

One of the major challenges in building a universal quantum computer is error correction. During any computation, errors are introduced by physical imperfections of the hardware. But while classical computers allow for simple error correction based on duplicating information, the no-cloning theorem of quantum mechanics requires more complex solutions. The most well-known proposal of \textit{surface codes} prescribes to encode one ``logical qubit'' into a topological state of several ``physical qubits''. Measurements on these physical qubits reveal a ``footprint'' of the chain of error events called a \textit{syndrome}. A \textit{decoder} maps a syndrome to an error sequence, which, once known, can be corrected by applying the same error sequence again, and without affecting the logical qubits that store the actual quantum information. Roughly stated, the art of quantum error correction is therefore to predict errors from a syndrome - a task that naturally fits the framework of machine learning.

In the past few years, various models have been applied to quantum error correction, ranging from supervised to unsupervised and reinforcement learning. The details of their application became increasingly complex. One of the first proposals deploys a Boltzmann machine trained by a data set of pairs $(\text{error}, \text{syndrome})$, which specifies the probability $p(\text{error}, \text{syndrome})$, which can be used to draw samples from the desired distribution $p(\text{error}| \text{syndrome})$ \citep{torlai_neural_2017}.  This simple recipe shows a performance for certain kinds of errors comparable to common benchmarks. The relation between syndromes and errors can likewise be learned by a feed-forward neural network \citep{varsamopoulos2017decoding, krastanov2017deep, maskara2019advantages}.  However, these strategies suffer from scalability issues, as the space of possible decoders explodes and data acquisition becomes an issue. More recently, neural networks have been combined with the concept of renormalization group to address this problem \citep{varsamopoulos2018designing}, and the significance of different hyper-parameters of the neural network has been studied \citep{varsamopoulos2019decoding}.

Besides scalability, an important problem in quantum error correction is that the syndrome measurement procedure could also introduce an error, since it involves applying a small quantum circuit. This setting increases the problem complexity but is essential for real applications. Noise in the identification of errors can be mitigated by doing repeated cycles of syndrome measurements. To consider the additional time dimension, recurrent neural network architectures have been proposed \citep{baireuther_2018}. Another avenue is to consider decoding as a reinforcement learning problem \citep{sweke2018reinforcement}, in which an agent can choose consecutive operations acting on physical qubits (as opposed to logical qubits) to correct for a syndrome and gets rewarded if the sequence corrected the error.

 While much of machine learning for error correction focuses on \textit{surface codes} that represent a logical qubit by physical qubits according to some set scheme, reinforcement agents can also be set up agnostic of the code (one could say they learn the code along with the decoding strategy). This has been done for quantum memories, a system in which quantum states are supposed to be stored rather than manipulated \citep{nautrup2018optimizing}, as well as in a feedback control framework which protects qubits against decoherence \citep{fosel_reinforcement_2018}. Finally, beyond traditional reinforcement learning, novel strategies such as \textit{projective simulation} can be used to combat noise \cite{tiersch2015adaptive}.

As a summary, machine learning for quantum error correction is a problem with several layers of complexity that, for realistic applications, requires rather complex learning frameworks. Nevertheless, it is a very natural candidate for machine learning, and especially reinforcement learning.


\section{Chemistry and Materials}
\label{sec:chem_mat}


Machine learning approaches have been applied to predict the energies and properties of molecules and solids, with the popularity of such applications increasing dramatically.
The quantum nature of atomic interactions makes energy evaluations computationally expensive, so ML methods are particularly useful when many such calculations are required.
In recent years, the ever-expanding applications of ML in chemistry and materials research include predicting the structures of related molecules, calculating energy surfaces based on molecular dynamics (MD) simulations, identifying structures that have desired material properties, and creating machine-learned density functionals.
For these types of problems, input descriptors must account for differences in atomic environments in a compact way.
Much of the current work using ML for atomistic modeling is based on early work describing the local atomic environment with symmetry functions for input into a atom-wise neural network \cite{Behler:2007kp}, representing atomic potentials using Gaussian process regression methods \cite{Bartok:2010js}, or using sorted interatomic distances weighted by the nuclear charge (the "Coulomb matrix") as a molecular descriptor \cite{Rupp:2012atomenergy}.
Continuing development of suitable structural representations is reviewed by \textcite{Behler:2016fh}.
A discussion of ML for chemical systems in general, including learning structure-property relationships, is found in the review by \textcite{Butler:2018fl}, with additional focus on data-enabled theoretical chemistry reviewed by \textcite{Rupp:2018kp}.
In the sections below, we present recent examples of ML applications in chemical physics.

\subsection{Energies and forces based on atomic environments}
\begin{figure*}[htb]
    \centering
    \includegraphics[width=\textwidth]{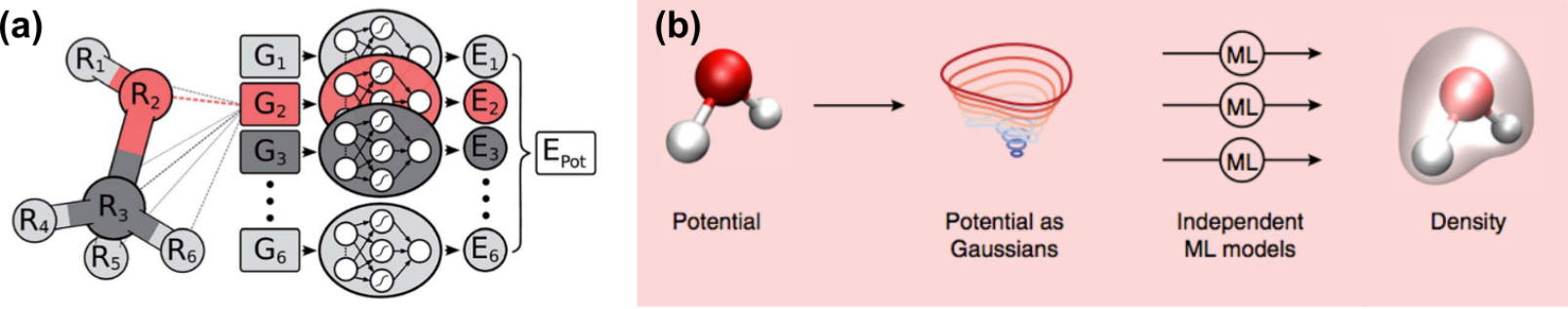}
    \caption{Several representations are currently used to describe molecular systems in ML models, including (a) atomic coordinates, with symmetry functions encoding local bonding environments, as inputs to element-based neural networks (Reproduced from ~\cite{Gastegger:2017bi}) and (b) nuclear potentials approximated by a sum of Gaussian functions as inputs kernel ridge regression models for electron densities (Modified from ~\cite{Brockherde:2017vd}).}
    \label{fig:molecule_reps}
\end{figure*}

One of the primary uses of ML in chemistry and materials research is to predict the relative energies for a series of related systems, most typically to compare different structures of the same atomic composition.
These applications aim to determine the structure(s) most likely to be observed experimentally or to identify molecules that may be synthesizable as drug candidates.
As examples of supervised learning, these ML methods employ various quantum chemistry calculations to label molecular representations ($X\mu$) with corresponding energies ($y\mu$)to generate the training (and test) data sets $\{X_\mu, y_\mu\}_{\mu=1}^n$.
For quantum chemistry applications, NN methods have had great success in predicting the relative energies of a wide range of systems, including constitutional isomers and non-equilibrium configurations of molecules, by using many-body symmetry functions that describe the local atomic neighborhood of each atom \cite{Behler:2016fh}.
Many successes in this area have been derived from this type of atom-wise decomposition of the molecular energy, with each element represented using a separate NN \cite{Behler:2007kp} (see Fig.~\ref{fig:molecule_reps}(a)).
For example, ANI-1 is a deep NN potential successfully trained to return the density functional theory (DFT) energies of any molecule with up to 8 heavy atoms (H, C, N, O) \cite{Smith:2017dq}.
In this work, atomic coordinates for the training set were selected using normal mode sampling to include some vibrational perturbations along with optimized geometries.
Another example of a general NN for molecular and atomic systems is the Deep Potential Molecular Dynamics (DPMD) method specifically created to run MD simulations after being trained on energies from bulk simulations \cite{Zhang:2018kz}.
Rather than simply include non-local interactions via the total energy of a system, another approach was inspired by the many-body expansion used in standard computational physics.
In this case adding layers to allow interactions between atom-centered NNs improved the molecular energy predictions \cite{Lubbers:2018in}.

The examples above use translation- and rotation-invariant representations of the atomic environments, thanks to the incorporation of symmetry functions in the NN input.
For some applications, such as describing molecular reactions and materials phase transformations, atomic representations must also be continuous and differentiable.
The smooth overlap of atomic positions (SOAP) kernels address all of these requirements by including a similarity metric between atomic environments \cite{bartok2013soap}.
Recent work to preserve symmetries in alternate molecular representations addresses this problem in different ways.
To capitalize on known molecular symmetries for "Coulomb matrix" inputs, both bonding (rigid) and dynamic symmetries have been incorporated to improve the coverage of training data in the configurational space \cite{Chmiela:2018ge}.
This work also includes forces in the training, allowing for MD simulations at the level of coupled cluster calculations for small molecules, which would traditionally be intractable.
Molecular symmetries can also be learned, as shown in determining local environment descriptors that make use of continuous-filter convolutions to describe atomic interactions \cite{Schutt:2018hm}.
Further development of atom environment descriptors that are compact, unique, and differentiable will certainly facilitate new uses for ML models in the study of molecules and materials.

However, machine learning has also been applied in ways that are more closely integrated with conventional approaches, so as to be more easily incorporated in existing codes.
For example, atomic charge assignments compatible with classical force fields can be learned, without the need to run a new quantum mechanical calculation for each new molecule of interest \cite{Sifain:2018fr}.
In addition, condensed phase simulations for molecular species require accurate intra- and intermolecular potentials, which can be difficult to parameterize.
To this end, local NN potentials can be combined with physically-motivated long-range Coulomb and van der Waals contributions to describe larger molecular systems \cite{Yao:2018kn}.
Local ML descriptions can also be successfully combined with many-body expansion methods to allow application of ML potentials to larger systems, as demonstrated for water clusters \cite{Nguyen:2018iy}.
Alternatively, intermolecular interactions can be fitted to a set of ML models trained on monomers to create a transferable model for dimers and clusters \cite{Bereau:2018ig}.

\subsection{Potential and free energy surfaces}
Machine learning methods are also employed to describe free energy surfaces. Rather than learning the potential energy of each molecular conformation directly as described above, an alternate approach is to learn the free energy surface of a system as a function of collective variables, such as global Steinhardt order parameters or a local dihedral angle for a set of atoms. A compact ML representation of a free energy surface (FES) using a NN allows improved sampling of the high dimensional space when calculating observables that depend on an ensemble of conformers. For example, a learned FES can be sampled to predict the isothermal compressibility of solid xenon under pressure, or the expected NMR spin-spin J couplings of a peptide \cite{Schneider:2017cj}. Small NN's representing a FES can also be trained iteratively using data points generated by on-the-fly adaptive sampling \cite{Sidky:2018ji}. This promising approach highlights the benefit of using a smooth representation of the full configurational space when using the ML models themselves to generate new training data. As the use of machine-learned FES representations increases, it will be important to determine the limit of accuracy for small NN's and how to use these models as a starting point for larger networks or other ML architectures.

Once the relevant minima have been identified on a FES, the next challenge is to understand the processes that take a system from one basin to another.
For example, developing a Markov state model to describe conformational changes requires dimensionality reduction to translate molecular coordinates into the global reaction coordinate space. To this end, the power of deep learning with time-lagged autoencoder methods has been harnessed to identify slowly changing collective variables in peptide folding examples \cite{Wehmeyer:2018bk}.
A variational NN-based approach has also been used to identify important kinetic processes during protein folding simulations and provides a framework for unifying coordinate transformations and FES surface exploration \cite{Mardt:2018bl}.
A promising alternate approach is to use ML to sample conformational distributions directly. Boltzmann generators can sample the equilibrium distribution of a collective variable space and subsequently provide a set of states that represent the distribution of states on the FES \cite{Noe:2019ki}.

Furthermore, the long history of finding relationships between minima on complex energy landscapes may also be useful as \emph{we} learn to understand why ML models exhibit such general success.
Relationships between the methods and ideas currently used to describe molecular systems and the corresponding  are reviewed in \cite{Ballard:2017eu}.
Going forward, the many tools developed by physicists to explore and quantify features on energy landscapes may be helpful in creating new algorithms to efficiently optimize model weights during training. (See also the related discussion in Sec.~\ref{sec:landcapes}.)
This area of interdisciplinary research promises to yield methods that will be useful in both machine learning and physics fields.

\subsection{Materials properties}
Using learned interatomic potentials based on local environments has also afforded improvement in the calculation of materials properties.
Matching experimental data typically requires sampling from the ensemble of possible configurations, which comes at a considerable cost when using large simulation cells and conventional methods.
Recently, the structure and material properties of amorphous silicon were predicted using molecular dynamics (MD) with a ML potential trained on density functional theory (DFT) calculations for only small simulation cells \cite{Deringer:2018in}.
Related applications of using ML potentials to model the phase change between crystalline and amorphous regions of materials such as GeTe and amorphous carbon are reviewed by \textcite{Sosso:2018hv}.
Generating a computationally-tractable potential that is sufficiently accurate to describe phase changes and the relative energies of defects on both an atomistic and material scale is quite difficult, however the recent success for silicon properties indicates that ML methods are up to the challenge \cite{Bartok:2018ih}.

Ideally, experimental measurements could also be incorporated in data-driven ML methods that aim to predict material properties.
However, reported results are too often limited to high-performance materials with no counter examples for the training process.
In addition, noisy data is coupled with a lack of precise structural information needed for input into the ML model.
For for organic molecular crystals, these challenges were overcome for predictions of NMR chemical shifts, which are very sensitive to local environments, by using a Gaussian process regression framework trained on DFT-calculated values of known structures \cite{Paruzzo:2018kl}. Matching calculated values with experimental results prior to training the ML model enabled the validation of a predicted pharmaceutical crystal structure.

\begin{figure*}[htb]
    \centering
    \includegraphics[width=\textwidth]{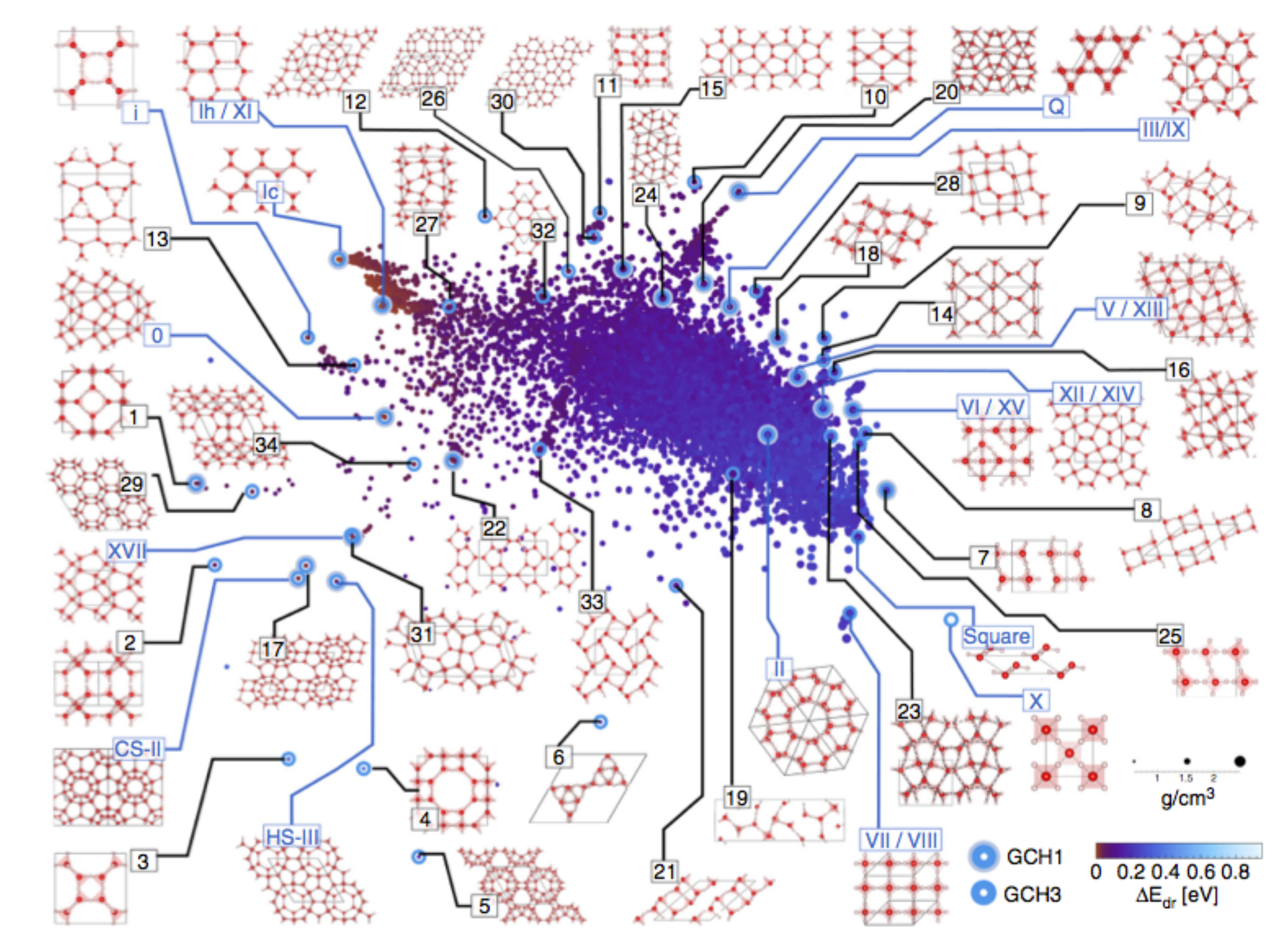}
    \caption{Clustering thousands of possible ice structures based on machine-learned descriptors identifies observed forms and groups similar structures together. Reproduced from ~\cite{Engel:2018ki}.}
    \label{fig:ice_structures}
\end{figure*}

Other intriguing directions include identification of structurally similar materials via clustering and using convex hull construction to determine which of the many predicted structures should be most stable under certain thermodynamic constraints \cite{Anelli:2018bw}.  Using kernel-PCA descriptors for the construction of the convex hull has been applied to identify crystalline ice phases and was shown to cluster thousands structures which differ only by proton disorder or stacking faults \cite{Engel:2018ki} (see Fig.~\ref{fig:ice_structures}).
Machine-learned methods based on a combination of supervised and unsupervised techniques certainly promises to be a fruitful research area in the future. In particular, it remains an exciting challenge to identify, predict, or even suggest materials that exhibit a particular desired property.

\subsection{Electron densities for density functional theory}
In many of the examples above, density functional theory calculations have been used as the source of training data.
It is fitting that machine learning is also playing a role in creating new density functionals.
Machine learning is a natural choice for situations such as DFT where we do not have knowledge of the functional form of an exact solution.
The benefit of this approach to identifying a density functional was illustrated by approximating the kinetic energy functional of an electron distribution in a 1D potential well \cite{Snyder:2012jm}.
For use in standard Kohn-Sham based DFT codes, the derivative of the ML functional must also be used to find the appropriate ground state electron distribution.  Using kernel ridge regression without further modification can lead to noisy derivatives, but projecting the resulting energies back onto the learned space using PCA resolves this issue \cite{Li:2015ej}.
A NN-based approach to learning the exchange-correlation potential has also been demonstrated for 1D systems \cite{Nagai:2018bk}. In this case, the ML method makes direct use of the derivatives generated during the NN training steps.

It is also possible to bypass the functional derivative entirely by using ML to generate the appropriate ground state electron density that corresponds to a nuclear potential \cite{Brockherde:2017vd}, as shown in Fig.~\ref{fig:molecule_reps}(b).
Furthermore, this work demonstrated that the energy of a molecular system can also be learned with electron densities as an input, enabling reactive MD simulations of proton transfer events based on DFT energies.
Intriguingly, an approximate electron density, such as a sum of densities from isolated atoms, has also been successfully employed as the input for predicting molecular energies \cite{Eickenberg:2018hta}.
A related approach for  periodic crystalline solids used local electron densities from an embedded atom method to train Bayesian ML models to return total system energies \cite{Schmidt:2018ix}.
Since the total energy is an extensive property, a scalable NN model based on summation of local electron densities has also been developed to run large DFT-based simulations for 2D porous graphene sheets \cite{Mills:2019ev}.
With these successes, it has become clear that given density functional, machine learning offers new ways to learn both the electron density and the corresponding system energy.

Many human-based approaches to improving the approximate functionals in use today rely on imposing physically-motivated constraints.
So far, including these types of restrictions on ML-based methods has met with only partial success.
For example, requiring that a ML functional fulfill more than one constraint, such as a scaling law and size-consistency, improves overall performance in a system-dependent manner \cite{Hollingsworth:2018er}.
Obtaining accurate derivatives, particularly for molecules with conformational changes, is still an open question for physics-informed ML functionals and potentials that have not been explicitly trained with this goal \cite{Snyder:2012jm, Bereau:2018ig}.

\subsection{Data set generation}
As for other applications of machine learning, comparison of various methods requires standardized data sets.
For quantum chemistry, these include the 134,000 molecules in the QM9 data set \cite{Ramakrishnan:2014fd} and the COMP6 benchmark data set composed of randomly-sampled subsets of other small molecule and peptide data sets, with each entry optimized using the same computational method \cite{Smith:2018ho}.

In chemistry and materials research, computational data are often expensive to generate, so selection of training data points must be carefully considered.
The input and output representations also inform the choice of data. Inspection of ML-predicted molecular energies for most of the QM9 data set showed the importance of choosing input data structures that convey conformer changes \cite{Faber:2017cs}.
In addition, dense sampling of the chemical composition space is not always necessary.
For example, the initial ANI training set of 20 million molecules could be replaced with 5.5 million training points selected using an active learning method that added poorly predicted molecular examples from each training cycle \cite{Smith:2018ho}.
Alternate sampling approaches can also be used to more efficiently build up a training set.
These range from active learning methods that estimate errors from multiple NN evaluations for new molecules\cite{Gastegger:2017bi} to generating new atomic configurations based on MD simulations using a previously-generated model \cite{zhang2018active}. Interesting, statistical-physics-based, insight into theoretical aspects of such active learning was presented in \cite{seung1992query}.

Further work in this area is needed to identify the atomic compositions and configurations that are most important to differentiating candidate structures.
While NN's have been shown to generate accurate energies, the amount of data required to prevent over-fitting can be prohibitively expensive in many cases.
For specific tasks, such as predicting the anharmonic contributions to vibrational frequencies of the small molecule formaldehye, Gaussian process methods were more accurate, and used fewer points than a NN, although these points need to be selected more carefully \cite{Kamath:2018gf}.
Balancing the computational cost of data generation, ease of model training, and model evaluation time continues to be an important consideration when choosing the appropriate ML method for each application.

\subsection{Outlook and Challenges}
Going forward, ML models will benefit from including methods and practices developed for other problems in physics.
While some of these ideas are already being explored, such as exploiting input data symmetries for molecular configurations, there are still many opportunities to improve model training efficiency and regularization.
Some of the more promising (and challenging) areas include applying methods for exploration of high-dimensional landscapes for parameter/hyper-parameter optimization and identifying how to include boundary behaviors or scaling laws in ML architectures and/or input data formats.
To connect more directly to experimental data, future physics-based ML methods should account for uncertainties and/or errors from calculations and measured properties to avoid over-fitting and improve transferability of the models.

\section{AI acceleration with classical and quantum hardware}
\label{sec:instruments}

There are areas where physics can contribute to machine learning by other means than tools for theoretical investigations and domain-specific problems. Novel hardware platforms may help with expensive information processing pipelines and extend the number crunching facilities of CPUs and GPUs. Such hardware-helpers are also known as ``AI accelerators'', and physics research has to offer a variety of devices that could potentially enhance machine learning.

\subsection{Beyond von Neumann architectures}

When we speak of computers, we usually think of universal digital computers based on electrical circuits and Boolean logic. This is the so-called "von Neumann" paradigm of modern computing. But any physical system can be interpreted as a way to process information, namely by mapping the input parameters of the experimental setup to measurement results, the output. This way of thinking is close to the idea of analog computing, which has been -- or so it seems \cite{lundberg2005history, ambs2010optical} -- dwarfed by its digital cousin for all but very few applications. In the context of machine learning however, where low-precision computations have to be executed over and over, analog and special-purpose computing devices have found a new surge of interest. The hardware can be used to emulate a full model, such as neural-network inspired chips \cite{ambrogio2018equivalent}, or it can outsource only a subroutine of a computation, as done by Field-Programmable Gate Arrays (FPGAs) and Application-Specific Integrated Circuits (ASICs) for fast linear algebra computations \cite{jouppi2017datacenter, markidis2018nvidia}.

In the following, we present selected examples from various research directions that investigate how hardware platforms from physics labs, such as optics, nanophotonics and quantum computers, can become novel kinds of AI accelerators.

\begin{figure*}
    \centering
    \includegraphics[width=\textwidth]{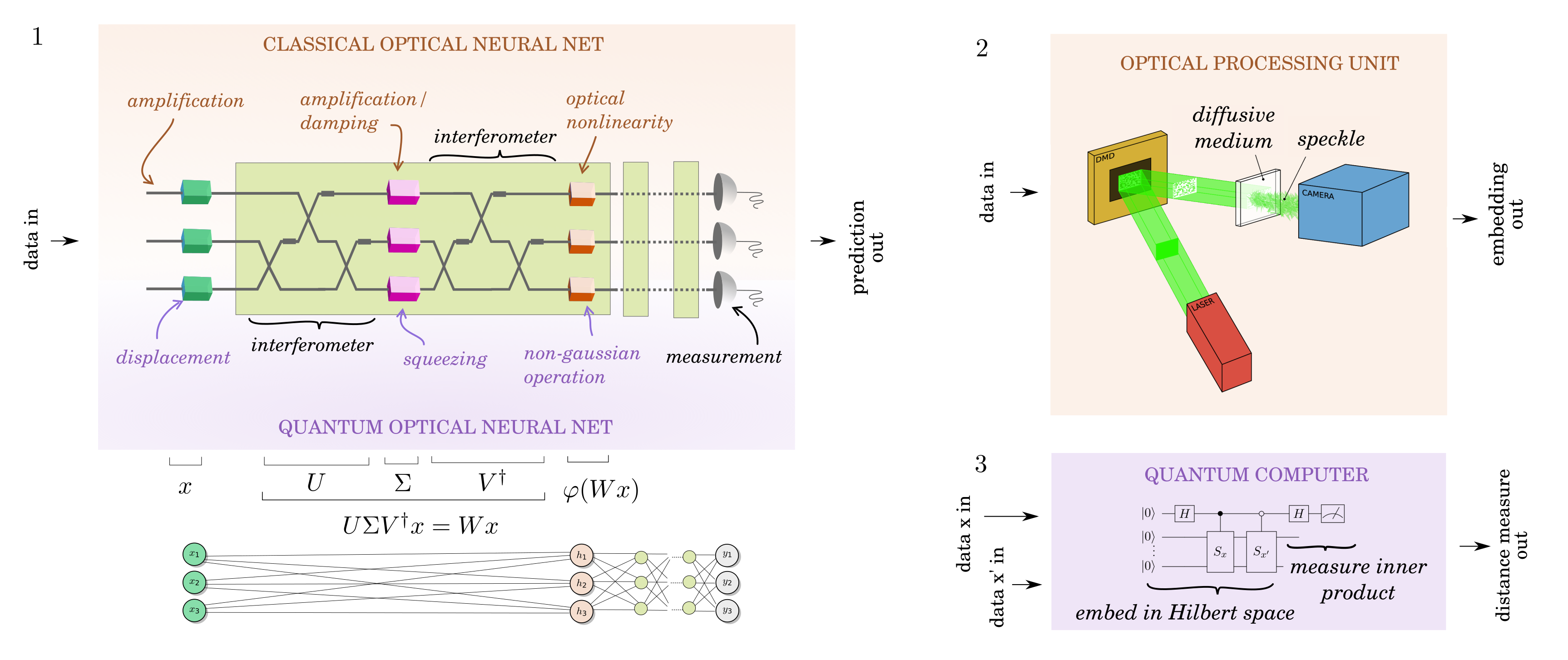}
    \caption{Illustrations of the methods discussed in the text. 1. Optical components such as interferometers and amplifiers can emulate a neural network that layer-wise maps an input $x$ to $\varphi(Wx)$ where $W$ is a learnable weight matrix and $\varphi$ a nonlinear activation. Using quantum optics components such as displacement and squeezing, one can encode information into quantum properties of light and turn the neural net into a universal quantum computer. 2. Random embedding with an Optical Processing Unit. Data is encoded into the laser beam through a spatial light modulator (here, a DMD), after which a diffusive medium generates the random features. 3. A quantum computer can be used to compute distances between data points, or ``quantum kernels''. The first part of the quantum algorithm uses routines $S_x$, $S_{x'}$ to embed the data in Hilbert space, while the second part reveals the inner product of the embedded vectors. This kernel can be further processed in standard kernel methods such as support vector machines.   }
    \label{fig:illustration}
\end{figure*}

\subsection{Neural networks running on light}

Processing information with optics is a natural and appealing alternative - or at least  complement - to all-silicon computers: it is fast, it can be made massively parallel, and requires very low power consumption. Optical interconnects are already widespread, to carry information on short or long distances, but light interference properties also can be leveraged in order to provide more advanced processing. In the case of machine learning there is one more perk. Some of the standard building blocks in optics labs have a striking resemblance with the way information is processed with neural networks \cite{shen2017deep, lin2018all, killoran2018continuous}, an insight that is by no means new \cite{lu1989two}. An example for both large bulk optics experiments and on-chip nanophotonics are networks of \textit{interferometers}. Interferometers are passive optical elements made up of beam splitters and phase shifters \cite{clements16, reck94}. If we consider the amplitudes of light modes as an incoming signal, the interferometer effectively applies a unitary transformation to the input (see Figure \ref{fig:illustration} left). Amplifying or damping the amplitudes can be understood as applying a diagonal matrix. Consequently, by means of a singular value decomposition, an amplifier sandwiched by two interferometers implements an arbitrary matrix multiplication on the data encoded into the optical amplitudes. Adding a non-linear operation -- which is usually the hardest to precisely control in the lab -- can turn the device into an emulator of a standard neural network layer \cite{shen2017deep, lin2018all}, but at the speed of light.

An interesting question to ask is: what if we use \textit{quantum} instead of classical light? For example, imagine the information is now encoded in the \textit{quadratures} of the electromagnetic field. The quadratures are - much like position and momentum of a quantum particle - two non-commuting operators that describe light as a quantum system. We now have to exchange the setup to quantum optics components such as squeezers and displacers, and get a neural network encoded in the \textit{quantum} properties of light \cite{killoran2018continuous}. But there is more: Using multiple layers, and choosing the `nonlinear operation' as a ``non-Gaussian'' component (such as an optical ``Kerr non-linearity'' which is admittedly still an experimental challenge), the optical setup becomes a \textit{universal quantum computer}. As such, it can run any computations a quantum computer can perform - a true \textit{quantum neural network}. There are other variations of quantum optical neural nets, for example when information is encoded into discrete rather than continuous-variable properties of light \cite{steinbrecher2018quantum}. Investigations into what these quantum devices mean for machine learning, for example whether there are patterns in data that can be easier recognized, have just begun.

\subsection{Revealing features in data}

One does not have to implement a full machine learning model on the physical hardware, but can outsource single components.  An example which we will highlight as a second application is data preprocessing or feature extraction. This includes mapping data to another space where it is either compressed or `blown up', in both cases revealing its features for machine learning algorithms.

One approach to data compression or expansion with physical devices leverages the very statistical nature of many machine learning algorithms. Multiple light scattering can generate the very high-dimensional randomness needed for so-called {\it random embeddings} (see Figure \ref{fig:illustration} top right). In a nutshell, the multiplication of a set of vectors by the same random matrix is approximately distance-preserving \cite{JL84}. This can be used for dimensionality reduction, i.e., data compression, in the spirit of {\it compressed sensing} \cite{CS06} or for efficient nearest neighbor search with locality sensitive hashing. This can also be used for dimensionality expansion, where in the limit of large dimension it approximates a well-defined {\it kernel} \cite{Kernel16}. Such devices can be built in free-space optics, with coherent laser sources, commercial light modulators and CMOS sensors, and a well-chosen scattering material (see Fig.\ref{fig:illustration} 2a). Machine learning applications range from  transfer learning for deep neural networks, time series analysis - with a feedback loop implementing so-called {\it echo-state networks} \cite{Dong18}, or change-point detection \cite{newma18}. For large-dimensional data, these devices already outperform CPUs or GPUs both in speed and power consumption.

\subsection{Quantum-enhanced machine learning}

 A fair amount of effort in the field of \textit{quantum machine learning}, a field that investigates intersections of quantum information and intelligent data mining \cite{biamonte17, schuld2018supervised}, goes into applications of near-term quantum hardware for learning tasks \cite{perdomo17a}. These so-called \textit{Noisy Intermediate-Scale Quantum} or `NISQ' devices are not only hoped to enhance machine learning applications in terms of speed, but may lead to entirely new algorithms inspired by quantum physics. We have already mentioned one such example above, a quantum neural network that can emulate a classical neural net, but go beyond. This model falls into a larger class of \textit{variational} or \textit{parametrized} quantum machine learning algorithms \cite{mcclean2016theory, mitarai2018quantum}. The idea is to make the quantum algorithm, and thereby the device implementing the quantum computing operations, depend on parameters $\theta$ that can be trained with data. Measurements on the ``trained device'' represent new outputs, such as artificially generated data samples of a generative model, or classifications of a supervised classifier.

 Another idea of how to use quantum computers to enhance learning is inspired by kernel methods \cite{hofmann08}  (see Figure \ref{fig:illustration} bottom right). By associating the parameters of a quantum algorithm with an input data sample $x$, one effectively embeds $x$ into a quantum state $|\psi(x)\rangle$ described by a vector in Hilbert space \cite{schuld18feat, havlicek2018}. A simple interference routine can measure overlaps between two quantum states prepared in this way. An overlap is an inner product of vectors in Hilbert space, which in the machine literature is known as a \textit{kernel}, a distance measure between two data points. As a result, quantum computers can compute rather exotic kernels that may be classically intractable, and it is an active area of research to find interesting quantum kernels for machine learning tasks.

 Beyond quantum kernels and variational circuits, quantum machine learning presents many other ideas that use quantum hardware as AI accelerators, for example as a sampler for training and inference in graphical models \cite{adachi15, benedetti2017quantum}, or for linear algebra computations \cite{lloyd14}\footnote{Many quantum machine learning algorithms based on linear algebra acceleration have recently been shown to make unfounded claims of exponential speedups \cite{tang2018quantum}, when compared against classical algorithms for analysing low-rank datasets with strong sampling access. However, they are still interesting in this context where even constant speedups make a difference.}.  Another interesting branch of research investigates how quantum devices can directly analyze the data produced by quantum experiments, without making the detour of measurements \cite{cong2018quantum}. In all these explorations, a major challenge is the still severe limitations in current-day NISQ devices which reduce numerical experiments on the hardware to proof-of-principle demonstrations, while theoretical analysis remains notoriously difficult in machine learning.

\subsection{Outlook and Challenges}
The above examples demonstrate a  way of how physics research can contribute to machine learning, namely by investigating new hardware platforms to execute tiresome computations. While standard von Neumann technologies struggle to keep pace with Moore's law, this opens a number of opportunities for novel computing paradigms. In their simplest embodiment, these take the form of specialized accelerator devices, plugged onto standard servers and accessed through custom APIs. Future research focuses on the scaling-up of such hardware capabilities, hardware-inspired innovation to machine learning, and adapted programming languages as well as compilers for the optimized distribution of computing tasks on these hybrid servers.

\section{Conclusions and Outlook}
\label{sec:outlook}

A number of overarching themes become apparent after reviewing the ways in which machine learning is used in or has enhanced the different disciplines of physics.
First of all, it is clear that the interest in machine learning techniques suddenly surged in recent years. This is true even in areas such as statistical physics and high-energy physics where the connection to machine learning techniques has a long history.
%
We are seeing the research move from an exploratory efforts on toy models  towards the use of real experimental data. We are also seeing an evolution in the understanding and limitations of these approaches and situations in which the performance can be justified theoretically. A healthy and critical engagement with the potential power and limitations of machine learning includes an analysis of where these methods break and what they are distinctly \textit{not} good at.

Physicist are notoriously hungry for very detailed understanding of why and when their methods work. As machine learning is incorporated into the physicist's toolbox, it is reasonable to expect that physicist may shed light on some of the notoriously difficult questions machine learning is facing. Specifically, physicists are already contributing to issues of interpretability, techniques to validate or guarantee the results, and principle ways to chose the various parameters of the neural networks architectures.

One direction in which the physics community has much to learn from the machine learning community is the culture and practice of sharing code and developing carefully-crafted, high-quality benchmark datasets. Furthermore, physics would do well to emulate the practices of developing user-friendly and portable implementations of the key methods, ideally with the involvement of professional software engineers.

The picture that emerges from the level of activity and the enthusiasm surrounding the first success stories is that the interaction between machine learning and the physical sciences is merely in its infancy, and we can anticipate more exciting results stemming from this interplay between machine learning and the physical sciences.



\section*{Acknowledgements}
This research was supported in part by the National Science Foundation under Grants Nos. NSF PHY-1748958, ACI-1450310, OAC-1836650, and DMR-1420073, the US Army Research Office under contract/grant number W911NF-13-1-0387, as well as the ERC under the European Unions Horizon 2020 Research and Innovation Programme Grant Agreement 714608-SMiLe. Additionally, we would like to thank the support of the Moore and Sloan foundations, the Kavli Institute of Theoretical Physics of UCSB, and the Institute for Advanced Study. Finally, we would like to thank Michele Ceriotti, Yoav Levine, Andrea Rocchetto, Miles Stoudenmire, and Ryan Sweke.

\bibliographystyle{apsrmp4-1}
\bibliography{biblio}

\end{document}